\documentclass[review]{elsarticle}

\usepackage{lineno,hyperref}
\usepackage{subfig}
\usepackage[]{graphicx}
\usepackage{color}
\usepackage{dcolumn}
\usepackage{amsmath}
\usepackage{amsfonts}
\usepackage{fullpage}
\usepackage{bm}
\usepackage{todonotes} 
\usepackage{algorithm}
\usepackage{algorithmicx,algpseudocode}
\usepackage{soul} 

\graphicspath{{Figures/}}

\usepackage{lipsum}
\makeatletter
 \def\ps@pprintTitle{%
 \let\@oddhead\@empty
 \let\@evenhead\@empty
 \def\@oddfoot{}%
 \let\@evenfoot\@oddfoot}
\makeatother

\begin{document}

\graphicspath{{Figures/}}

\begin{frontmatter}

\title{A unified framework for heat and mass transport at the atomic scale}

\author{Mauricio Ponga$^*$}
\address{Department of Mechanical Engineering, University of British Columbia, 2054 - 6250 Applied Science Lane, Vancouver, BC, V6T 1Z4, Canada}
\cortext[mycorrespondingauthor]{Corresponding author}
\ead{mponga@mech.ubc.ca}

\author{Dingyi Sun}
\address{School of Engineering, Brown University, 182 Hope St., Box D, Providence, RI 02912, United States}
\ead{dingyi\_sun@brown.edu}

\begin{abstract}

We present a unified framework to simulate heat and mass transport in systems of particles. The proposed framework is based on kinematic mean field theory and uses a phenomenological master equation to compute effective transport rates between particles without the need to evaluate operators. We exploit this advantage and apply the model to simulate transport phenomena at the nanoscale. We demonstrate that, when calibrated to experimentally-measured transport coefficients, the model can accurately predict transient and steady state temperature and concentration profiles even in scenarios where the length of the device is comparable to the mean free path of the carriers. Through several example applications, we demonstrate the validity of our model for all classes of materials, including ones that, until now, would have been outside the domain of computational feasibility.
\end{abstract}


\end{frontmatter}


{\color{blue} Nanoscale heat conduction is a subject of great interest due to its applications to the next-generation of nano- and micro-electronic devices, where the heat flux generated can be exceedingly large in comparison with that seen in the current generation of electronics \cite{Cahill:2003,NanoscaleEnergy:2005}. Thus, it is of utmost importance to understand how heat is carried at these small scales.} However, modeling {\color{blue} and simulation of} heat transport at the nanoscale is a complicated undertaking; the classical Fourier equation is no longer valid and common atomistic simulation techniques -- such as molecular dynamics (MD) -- are not able to model all classes of materials accurately. This is due to the fact that when the lengths of these devices become comparable to the mean free-path, the classical Fourier equation (FE) is no longer valid for predicting their behavior due to the fact that the heat carriers can scatter upon interaction with interfaces and defects, resulting in a lower conductivity than bulk materials \cite{Cahill:2003,NanoscaleEnergy:2005,Pop:2010}. As such, nanoscale thermal properties are intimately coupled to the distribution and evolution of defects, necessitating the development of new models that can accurately predict such nanoscale thermo-mechanical behavior.

To model transport of heat carriers and their interaction with defects, a typical approach is to employ molecular dynamics methods; this is acceptable as long as these carriers consist \emph{only} of phonons. Unfortunately, in most materials, heat is carried out by \emph{both} phonons and electrons; this is problematic because MD models do not account for heat carried by electrons. Thus, MD is only capable of accurate predictions for insulators, but not metallic materials. An approach that accurately accounts for transport of both phonons and electrons is based on the Boltzmann transport equation (BTE) \cite{Joshi:1993,Jou:1999,Chen:2001}; however, such approaches {\color{blue} are not amenable to} concurrent thermo-mechanical simulation of materials since the resolution of the BTE is exceedingly expensive and difficult, especially for fully three-dimensional problems.

To remedy this, we seek to develop a unified framework to simulate coupled transport phenomena including heat and mass transport problems at the atomic scale. Our goals are two fold. First, we wish to embed the transport model with atomic-scale techniques, such as MD or \emph{ab-initio} techniques to simulate coupled, thermo-chemo-mechanical problems at the nanoscale. Second, we seek to develop a universal formulation for both heat and mass transport such that relevant transport coefficients can be obtained from either experiments or atomic-scale simulations. Attaining these goals will allow for the study of coupled simulations of nanoscale materials and structures while keeping a notion of the mesh-free nature of MD, eliminating the need for regular meshes of basis sets.

In this work, we propose a new model for heat conduction based on a linearized Fokker-Planck equation. We formulate an empirical kinematic law for heat transport based on kinematic mean-field theory and transition state theory \cite{Martin:1990}. The result is a master equation which is analogous to equations governing nanoscale mass transport that have been demonstrated previously in the community \cite{CurtinDiff:2008,DMD-Li:2011,DMD-Dontsova:2014}. A key difference between our model and Fourier model is that atomic level information on the kinetic energy acts as the thermodynamic driving force for heat transport; in Fourier, transport is driven by temperature gradients. Furthermore, the proposed model is local, but the heat propagation has a finite velocity in contrast to the infinite propagation velocity of the Fourier model. Thus, our model offers an alternative approach to both the BTE and the FE that 1) can be calibrated to experimental measures, and 2) can be seamlessly coupled to different atomic-scale thermo-mechanical formulations. These features give our model unprecedented predictive capabilities. To fully realize this model in a computational environment, we couple our new model with the HotQC method \cite{Kulkarni:2008,Ariza:2012,Venturini:2014,Ponga:2015,Ponga:2016,Romero:2016,Ponga:2017} in order to simulate both nanoscale heat \emph{and} mass diffusion; we then demonstrate its applicability in various scenarios.

\section{Heuristic of the model}

{\color{blue} We now present the basis of the framework, which is partially inspired by kinematic mean field theory and transition state theory  \cite{Martin:1990}}. Consider a system of $N$ interacting particles and suppose that the system is discretized over a set of different sites. Introduce a probability density of the $i^\text{th}$ site in this system, which we will denote $f_i\in[0,1]$; this probability density could be associated to different fields of interest (e.g. \emph{local} atomic temperature, normalized kinetic energy of the particle or atomic molar fraction). We postulate a so-called \emph{master equation} which dictates the time evolution of probability densities,
\begin{equation}\label{eq:MasterEquation}
\dfrac{\partial f_i(t)}{\partial t}=\sum_{{\substack{ j =1 \\ j \neq i}}}^{N_n} \zeta_{ij}\left\{[f_j(t)(1-f_i(t)) \Gamma_{j \rightarrow i}(t)]-[f_i(t)(1-f_j(t))\Gamma_{i \rightarrow j}(t)] \right\},
\end{equation}
with $\zeta_{ij}$ being some {\color{blue} pair-wise exchange rate coefficient between two nearest sites. We will see that $\zeta_{ij}$  can be related to macroscopically-relevant material properties, such as the thermal diffusivity ($\alpha$) or the diffusivity  of different solute atoms and vacancies ($D_m$)}. The term $\Gamma_{i \rightarrow j}$ indicates the probability that the quantity $f(t)$ would be transported from the $i^{\text{th}}$ to the $j^{\text{th}}$ site. The asymmetry of two terms on the right hand side of Eq. \ref{eq:MasterEquation} suggests that there is a probability that a state $i$ of high energy would \emph{jump} to state $j$ of lower energy, and vice versa. As one would expect, however, the probability that a low energy state jumps to a high energy state should be very small. Finally, we note that the sum in Eq. \ref{eq:MasterEquation} is carried out over the nearest neighbor of the $i^{\text{th}}$ site, thus making our model \emph{local}.

We make a few additional remarks about the probability density. The steric factors $f_i(t)(1-f_j(t))$ denote that there is a \emph{maximum} probability that is allowed in all sites, i.e. $f_i^{\text{max}} = 1$, at each time step. In our model, we will see that this is equivalent to the notion of a maximum energy level or molar occupation. However, this is a relative value that can change with time. Therefore, we do not introduce any artificial limitations to the evolution of the energy in the system. 

{\color{blue} Let us know analyze the physical significance of the pair-wise exchange rate coefficient $\zeta_{ij}$ used in the \emph{master equation}. The factor $\zeta_{ij}$ quantifies the number of exchange events that occur between sites $i$ and $j$ per unit of time. Thus, $\zeta_{ij}$ has units of (time)$^{-1}$, as required. We see that while the net amount of exchange is controlled by the local energy levels between sites which are directly used to compute the probabilities $\Gamma_{i \rightarrow j}$, the rate of exchange is governed by the pair-wise exchange rate coefficient $\zeta_{ij}$. We will see in Section \ref{HeatConduction} that  this coefficient is linked to physically-relevant quantities, such as the thermal diffusivity or the mass diffusivity of vacancy or solutes. For instance, when simulating heat conduction, the pair-wise exchange rate coefficient $\zeta_{ij}$ will be larger for metals than for insulators}. 

Notice that Eq. \ref{eq:MasterEquation} is an \emph{empirical} law used to simulate diffusive transport phenomena in many situations. Thus, we adopt it as the main tool to simulate both heat and mass transport at the nanoscale. While mass transport has been successfully modeled with a form of the master equation in atomic scale systems, we present, for the first time, an analogous heat transport model based on Eq. \ref{eq:MasterEquation}. 

We close this section by pointing out that when the size of nanodevices is reduced such that is comparable with the mean free-path of the heat carriers, the deviations from classical observations become more obvious. When the device length is comparable with the mean free-path, the heat carriers move ballistically. The correct characterization of such transient state is extremely challenging and several approaches have been proposed \cite{Joshi:1993,Chen:2001}. In the current context, the proposed \emph{master equation} (Eq. \ref{eq:MasterEquation}) models diffusive phenomena and does not include memory and non-local effects present in devices with lengths comparable to the mean-free path. We also notice that these effects can be taken into account by expanding the \emph{master equation} Eq. \ref{eq:MasterEquation} to include a relaxation time. For instance, let us now evaluate the master equation at a time $t+\tau_r$, where $\tau_r$ is a characteristic relaxation time scale. If one expands the left hand side of Eq. \ref{eq:MasterEquation} we obtain a hyperbolic heat equation model, i.e., 
 \begin{equation} \label{eq:MasterEquationHyperbolic}
\dfrac{\partial f_i(t)}{\partial t} + \tau_r \dfrac{\partial^2 f_i(t)}{\partial t^2} =\sum_{{\substack{ j =1 \\ j \neq i}}}^{N_n} \zeta_{ij}\left\{[f_j(t)(1-f_i(t)) \Gamma_{j \rightarrow i}(t)]-[f_i(t)(1-f_j(t))\Gamma_{i \rightarrow j}(t)] \right\}.
\end{equation}
We notice that the modified master equation (Eq. \ref{eq:MasterEquationHyperbolic}) now includes memory effects and can reproduce wave-like behavior by including a relaxation time scale. In our model, we include size effects by using a size-dependent thermal conductivity term, as proposed by Alvarez et al. \cite{Alvarez:2007}. We will show in section \ref{Validation} that the proposed model is as good as the Fourier equation and is capable of predicting steady state solutions provided we have sensible boundary conditions. {\color{blue} Finally, unless otherwise specified, the relaxation time is taken to be zero, obtaining the diffusive regime. }

\section{Heat conduction model at the atomic scale} \label{HeatConduction}

We begin by taking an appropriate discretization of the domain. We let our sites coincide with the atomic positions of systems of interest. No restrictions are imposed on this discretization, thus allowing simulation of even particles in random positions, as is the case of most glassy materials.

We now assume that each particle has a \emph{local atomic temperature}, $T_i$, i.e., that this field can vary between atoms. We then introduce a normalized kinetic energy per site, $\theta_i = \frac{T_i -T_c}{T_h -T_c}$, with $T_h$ and $T_c$ being the maximum and minimum temperatures allowed in the system, respectively. Recall that this field is allowed to change at different time steps, thus accommodating heating effects and other scenarios. Normalization maps the temperature field to $\theta_i\in[0,1]$; this allows for interpretation of the normalized temperature as the probability of the $i^\text{th}$ site having a certain amount of kinetic energy. We further impose the condition that the heat carriers --- both phonons and electrons --- travel from site to site and seek effective transport between adjacent sites. 

Assume the energy exchange between two sites can be approximated by transition state theory. The probability that a heat carrier will travel from the site $i$ to site $j$ is given by $\Gamma_{i \rightarrow j} = \exp(\Delta e_{ij})$, where 
{\color{blue}
\begin{equation}
\Delta e_{ij} = \frac{-(e_i - e_j)}{\overline{e}_{ij}} = \left( -\frac{2(T_i - T_j)}{(T_{i}+T_{j})} \right)
\end{equation}}
represents the normalized kinetic energy difference between the sites, {\color{blue} $e_i = k_B T_i$,  $\overline{e}_{ij} =  \frac{k_B}{2} (T_i+T_j)$ ,and $k_B$ is the Boltzmann constant}.

With these considerations and some inspiration from kinematic mean field theory {\color{blue} \cite{Martin:1990}}, we now propose an \emph{empirical} kinematic law for heat transport at the atomic level,
\begin{equation} \label{eq:HeatTransport}
\frac{\partial \theta_i}{\partial t} = \sum_{{{\substack{ j =1 \\ j \neq i}}}}^{N_n} K_{ij} \lbrace \theta_j (1-\theta_i) \exp [\Delta e_{ji} ]  - \theta_i (1-\theta_j) \exp [ \Delta e_{ij}]  \rbrace,
\end{equation}
with $K_{ij}$ {\color{blue} being a pair-wise exchange rate thermal coefficient} and $\Delta e_{ij}$ being a thermodynamic driving force, as described before. Note that the sum in Eq. \ref{eq:HeatTransport} is arbitrarily chosen to be carried out over the nearest neighbors of the $i^{\text{th}}$ site.

We now seek to determine the {\color{blue} pair-wise exchange rate thermal coefficient} $K_{ij}$. This parameter {\color{blue} quantifies the number of thermal exchange events per of unit time between two sites, controlling the rate exchange in the master equation. $K_{ij}$ can be linked to} intrinsic properties of the material that depends on many factors (such as the group velocity, length of the device, impurities, and global temperature, frequency of the carriers, among others). Carrying out an asymptotic expansion of Eq. \ref{eq:HeatTransport} and assuming small temperature gradients, we can link $K_{ij}$ to experimentally measured thermal diffusivity, $\alpha(L)$. We find this relation to be

\begin{equation}
K_{ij} = \frac{2 \alpha d}{Zb^2} = \frac{2 \lambda(L)  d}{ \rho C_p Zb^2}
\end{equation}
where $\lambda(L)$ is a length-dependent thermal conductivity used to account for size effects in nanodevices and nanowires, $\rho$ is the density of the material, $C_p$ is the specific heat at constant pressure of the material, $Z$ is the coordination number, $b$ is the Burgers vector of the material, and $d$ is the dimension of the problem. Following the works of Alvarez and Jou \cite{Alvarez:2007,Alvarez:2010}, the thermal conductivity and the device length are related through
\begin{equation} \label{eq:Size-conductivity}
\lambda(L)= \frac{\lambda_0 L^2}{2 \pi^2 \ell^2} \left[ \sqrt{1 + 4 \left( \frac{\pi \ell}{L} \right)^2}  -1 \right]
\end{equation}
where $\lambda_0$ is the bulk thermal conductivity. The ratio Kn $= \frac{\ell}{L}$ is usually called the Knudsen number, where $\ell$ is the mean-free path of the heat carriers and $L$ is the device length. 

\subsection{Thermodynamic properties of the proposed heat model}

Let us now analyze the energy balance and the entropy generation rate of Eq. \ref{eq:HeatTransport} and check that our proposed model satisfies the first and second law of thermodynamics. Consider the energy rate for the $i^\text{th}$ site when is interacting with its neighbors. The balance of energy at the $i^{\text{th}}$ site can be expressed as
\begin{equation}
\dot{e}_i = r_i + q_i
\end{equation}
where $\dot{e}_i $\footnote{{\color{blue} We note that the symbol $\dot{e}_i$ has the classical connotation of time derivative of the quantity $e_i$}.} is the rate of change of the energy of the i$^{\text{th}}$ site, $r_i$ is the net heat flux at the $i^\text{th}$ site due to the energy exchange between neighboring sites and $q_i$ is an \emph{internal} heat generation rate. $q_i$ is introduced to account for thermal heating when defects appear in the material. The net heat flux $r_i$ can be expressed as
\begin{equation}
 r_i = \sum_{\substack{\langle i, j \rangle \\ j \neq i}} R_{ij} = k_B \dot{T}_i,
\end{equation}
where $R_{ij}$ is the heat flux from the $j^\text{th}$ to the $i^\text{th}$ site and has units of energy per time, i.e., eV/sec. Following traditional conventions, we take flow \emph{into} the $i^\text{th}$ site to be positive and negative otherwise. With these definitions, let us now verify that Eq. \ref{eq:HeatTransport} obeys the fundamental laws of thermodynamics. To check for satisfaction of the first law of thermodynamics, we note that the steric factors $\theta_j (1-\theta_i)$ are symmetric if one permutes the index $i$ to $j$, ensuring heat flux from site $i$ to $j$ satisfies $R_{ij} = - R_{ji}$. 

For satisfaction of the second law of thermodynamics, let us consider two interacting sites and analyze the entropy generation rate at the $i^\text{th}$ site, 
\begin{equation}
 \dot{s}_i = \frac{\dot{e}_i}{T_i}
\end{equation}
Following Venturini \textit{et al.} \cite{Venturini:2014} and neglecting internal heat generation, the entropy generation rate between two interacting sites can be defined as
\begin{equation} \label{Clausius-DuhemInequality}
\sum_{ij} =  \frac{R_{ij}}{T_i} + \frac{R_{ji}}{T_j}.
\end{equation}
We then postulate the entropy generation rate $\sum_{ij}[\cdot] \ge 0$ to emulate the Clausius-Duhem inequality. By looking at the entropy generation rate and Eq. \ref{eq:HeatTransport}, we notice that heat flows from particles with high temperature to particles with low temperature, thus satisfying the discrete Clausius-Duhem inequality (Eq. \ref{Clausius-DuhemInequality}).

For a given, non-homogeneous, initial temperature field $\{ T \} = T|_{i=1}^N$, Eq. \ref{eq:HeatTransport} can be integrated to predict the evolution of the temperature field. Contrary to the Fourier equation, where the driving forces for heat conduction are given by temperature gradients, our proposed model (Eq. \ref{eq:HeatTransport}) uses the kinetic energy difference between sites as thermodynamic driving forces and allows for direct simulation of  diffusive heat conduction at the atomic-scale. Consequently, our model enables the prediction of heat conduction using information attained from techniques such as MD and/or \emph{ab-initio} methods. 

{\color{blue} 
\subsection{Extension to materials with anisotropic thermal conductivities}
So far, the thermal conductivity has been taken to be isotropic and a function of the device length. However, many materials show anisotropic properties, such as anisotropic thermal conductivities and diffusivities. In the presence of anisotropy, it is more convenient to express the thermal conductivity as a \emph{tensor} quantity. Let us now consider a possible extension of the model to account for anisotropy in the thermal diffusivity. For nanoscale devices, two directions are important to consider; the longitudinal and transversal directions with respect to the heat flux. Thermal conductivity can be quite anisotropic due to different device lengths in these directions. Assume the thermal conductivity tensor can be expressed as 

\begin{equation}
\boldsymbol{\lambda} = \left[
{\begin{array}{ccc}
 \lambda_1(L_x) & 0 & 0  \\
0 & \lambda_2(L_y) & 0  \\
0 & 0 & \lambda_3(L_z) \\
\end{array} }  \right].
\end{equation}
Here, the indexes $x,y$, and $z$ refer to the \emph{principal} directions of the thermal conductivity tensor. Anisotropic effects can be taken into account by using an effective thermal conductivity that depends on the relative position of the unit vector $\frac{{\bf r}_{ij}}{|{\bf r}_{ij}|}$, where ${\bf r}_{ij} = {\bf r}_{i} -{\bf r}_{j} $ is the relative distance between sites $i$ and $j$. Then, the pair-wise exchange rate thermal coefficient can be computed as 
\begin{equation}
K_{ij} = \frac{2 \lambda_{eff}(L,{\bf r}_{ij})  d}{ \rho C_p Zb^2} ,
\end{equation}
where $\lambda_{eff}(L,{\bf r}_{ij}) = \left[ \lambda_1(L_x) \alpha_1^2 + \lambda_2(L_y) \alpha_2^2 + \lambda_3(L_z) \alpha_3^2 \right]$, and $\alpha_k$ are the direction cosines of the vector ${\bf r}_{ij}$. $\lambda_{eff}(L,{\bf r}_{ij})$ is the \emph{effective} thermal conductivity that depends on the orientation between sites $i$ and $j$; the definition of this is orientation-dependent can be easily implemented in the code. We have done so and computed the heat conduction as an example that is described in Section \ref{AnisotropicModelValidation}. 
}

\subsection{Analogy with mass transport model}

As mentioned earlier, Eq. \ref{eq:HeatTransport} is only new in the context of heat conduction; an analogous master equation has been used previously to describe mass diffusion in atomic systems \cite{CurtinDiff:2008,DMD-Li:2011,DMD-Dontsova:2014}. Adjusting our fractional quantity of interest in Eq. \ref{eq:MasterEquation} to the atomic molar fraction of the $i^\text{th}$ site, $x_i$, we arrive at the governing equation for mass transport,
\begin{equation}\label{eq:MassTransport}
\frac{\partial  x_{i}}{\partial t}  =  \sum_{{{\substack{ j =1 \\ j \neq i}}}}^{N_{n}} D_{ij}[ x_{j} (1-x_{i}) \exp( \beta_j \Delta { \mu}_{ji}  ) - x_{i} (1-x_{j}) \exp( \beta_i \Delta { \mu}_{ij} ) ],
\end{equation}
where {\color[rgb]{0,0,1} $\beta_i = \frac{1}{k_B T_i}$ is the typical thermodynamic factor}, $D_{ij} = \nu_0 \exp (-\beta_i Q_m )$ {\color{blue} being a pair-wise exchange rate mass coefficient}, $\nu_0$ is an attempt frequency, and $Q_m$ is an energy barrier that the atoms need to overcome in order to hop from one site to another.  $\Delta \mu_{ij}=\mu_{i}-\mu_{j}$ is the difference in chemical potential and acts as the driving force for mass diffusion --- which is analogous to the difference in kinetic energy for heat conduction --- and $\mu_{i}=\partial\mathcal{F}/\partial x_{i}$ is the gradient of the free-energy, $\mathcal{F}$, with respect to the atomic molar fractions often called chemical potential. 

We are interested in adopting some of the methodology of HotQC \cite{Kulkarni:2008,Ariza:2012,Venturini:2014,Ponga:2015} to facilitate implementation. In short, the HotQC method uses the maximum-entropy principle to obtain the least biased probability distribution function in terms of the information-theoretical notion of entropy \cite{Jaynes:1957}, from which we obtain the grand canonical free-energy, Eq. \ref{Grand-Canonical-Free-Energy}. Using the HotQC formulation, the free-energy of the system can be written as 
\begin{equation} \label{Grand-Canonical-Free-Energy}
\mathcal{F} = k_B \sum_{i=1}^N \left[ \beta_i  \langle h_i \rangle_0 -3 +3 \log (\hbar \beta_i \omega_i) + x_{i} \log x_{i} \right],
\end{equation}
where $\hbar$ is the reduced Planck's constant, and $\omega_i$ is a vibrational atomic frequency.  The term $\langle h_i \rangle_0 = \frac{3}{2 \beta_i} + \langle V_i \rangle_0$ is the phase-averaged Hamiltonian of $i^{\text{th}}$ site and $\langle V_i \rangle_0$ represents a phase average of the interatomic interactions over the probability density function of the system. 

Following previous works \cite{CurtinDiff:2008,DMD-Li:2011,DMD-Dontsova:2014}, the associated diffusivity is
\begin{equation} \label{Pairwise-Diffusivity}
D_{ij} = \frac{2d D_m }{Z b^2}.
\end{equation}
where $D_m$ is the atom/vacancy diffusivity and can be computed as $D_m = \nu_0 \exp (-\beta_i Q_m )$. We now have the ability to directly calibrate diffusivity to experimental values or values from \emph{ab-initio} simulations. We then use the quasi-static version of HotQC alluded to earlier to bridge our heat conduction model to non-equilibrium statistical mechanics. In the remainder of this work, we will highlight several representative examples that we studied with our implementation. 

\section{Validation} \label{Validation}

\begin{figure}
\centering
\subfloat[]{\includegraphics[width=0.45\textwidth]{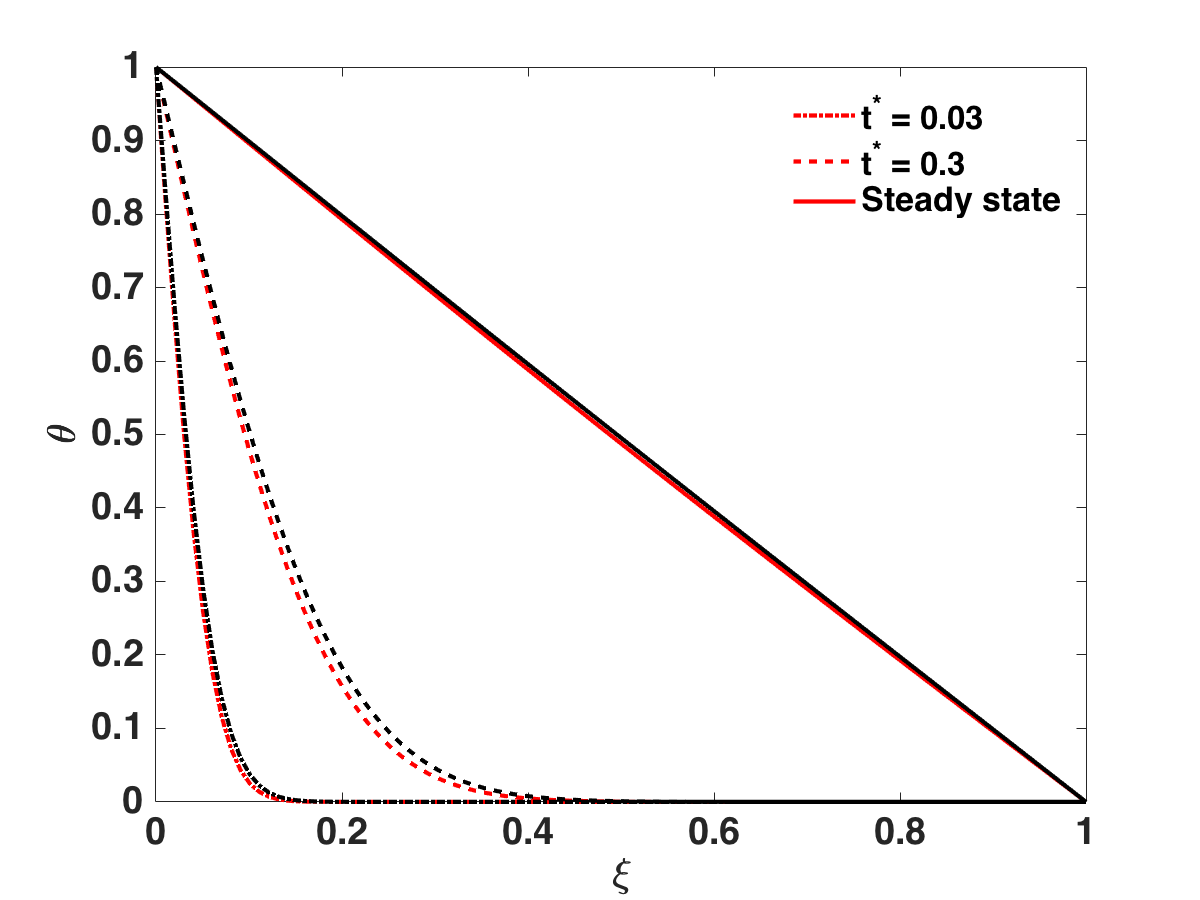}} 
\subfloat[]{\includegraphics[width=0.45\textwidth]{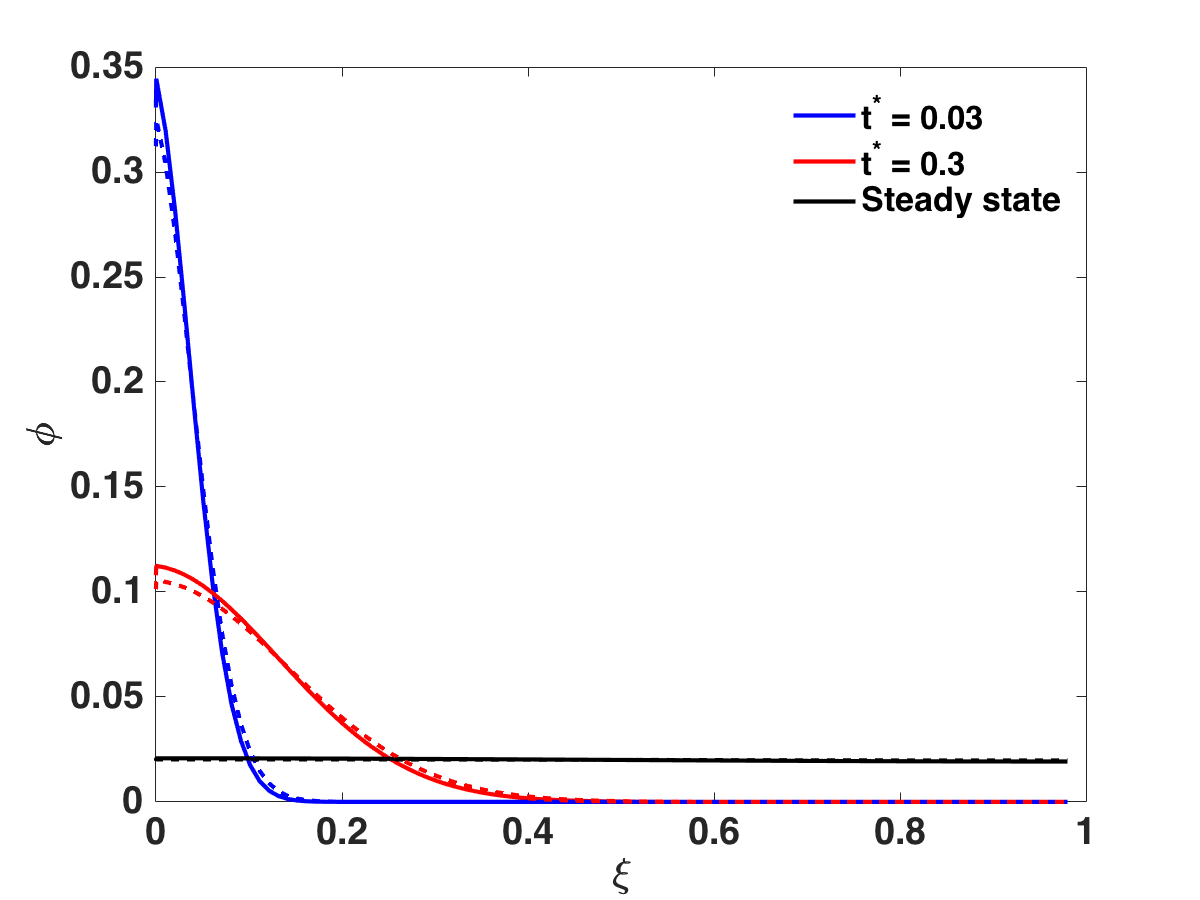}}
\caption{{\color[rgb]{0,0,1} a)} Time evolution of temperature along the domain for different times obtained with the Fourier equation (FE) (red) and the proposed method (black) when Kn $ \ll 1$. b) Heat flux in the sample for different times with FE (solid lines) and proposed approach (dashed lines).}
\label{Thermal-1}
\end{figure}

Let us now validate the proposed heat conduction model against the extended Fourier equation with the size-dependent thermal conductivity. This is essential to understanding the ability of the model to predict non-equilibrium properties. We start our analysis by considering a one dimensional chain of non-interacting atoms in their equilibrium position. We computed the evolution of the system using the classical Fourier law given by
\begin{equation} \label{FourierLaw}
{\bf J}_H= -\lambda(L)\nabla T
\end{equation}
and the classical heat equation,
\begin{equation} \label{ClassicalHeatConduction}
\dfrac{\partial T}{\partial t} = \alpha \nabla^2 T,
\end{equation}
where ${\bf J}_H$ is the heat flux. To non-dimensionalize our equations, we took $\lambda = \frac{1}{3} C_p v \ell$, $\tau = \frac{\ell}{v}$, where $\ell$ is the mean free path and $v$ is the speed of sound and $\tau$ is called the mean free time. $\Delta T = T_h - T_c$, $\xi = x/L$, $t^* = t/\tau$, $\theta = \frac{T -T_c}{\Delta T}$, ${\boldsymbol \phi } = \frac{{\bf J}_H}{C_p v \Delta T}$. Kn = $\ell /L$ is the Knudsen number.

Let us now examine the results of our implementation. We solved the heat conduction problem with the FE (Eq. \ref{ClassicalHeatConduction}) and compared against our new model with Dirichlet  boundary conditions, {\color{blue} i.e., $\theta(\xi=0,t^*)= 1$ and $\theta(\xi=1,t^*)= 0$)}. The classical approach was solved by discretizing the domain and computing the Laplace operator with finite differences (FD). {\color{blue} The FD implementation was carried out in MATLAB$^\circledR$ with a custom code where the Laplace operator is implemented with the second order central FD operator.} In both cases, the FD points were considered lattice positions and the mesh was the same for both methods. The temperature was integrated using an Euler forward algorithm; this led to a critical time step for integration of the heat equation of $t_c = \frac{b^2}{\alpha}$.

Figure \ref{Thermal-1} shows the time evolution of the temperature and the heat flux for a sample with Kn $\ll 1$ and $\lambda(L) = \lambda_0$, corresponding to the continuum domain where the Fourier equation is valid. The results obtained with the FE are shown in red while the results with the new model are shown in black. The agreement for this example is very good and only small differences are observed do to numeric differences. Comparisons with {\color{blue} Neumann} boundary conditions lead to same agreement with the FE and are omitted here.

\begin{figure}
\centering
\includegraphics[width=0.45\textwidth]{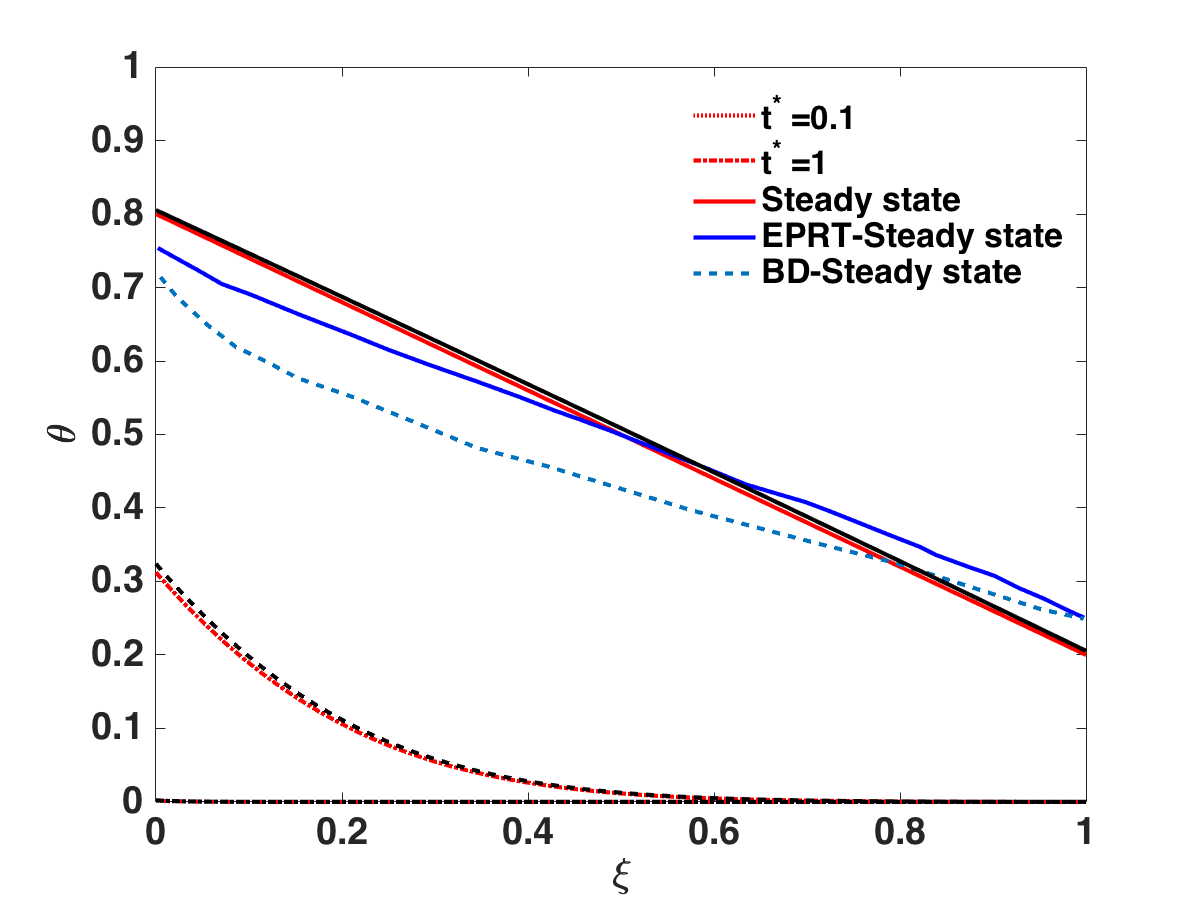}
\caption{Time evolution of temperature along the domain for different times obtained with the Fourier equation (FE) (red) the proposed method (black) and the EPRT model (blue) when Kn = 1. }
\label{Thermal-2}
\end{figure}

Next, we computed the time evolution of the temperature when Kn $= 1$. In this case, the mean free path is comparable with the device length; therefore, non-classical effects arise. Our intention is to assess the ability of the FE and the proposed model to capture non-classical effects and compare them with more sophisticated techniques, such as the equation of phonon radiative transfer (EPRT) \cite{Joshi:1993} and the ballistic diffusive BD \cite{Chen:2001} approaches {\color{blue} where the BTE is solved}. To make a fair comparison, we followed the work of Alvarez \emph{et al.} \cite{Alvarez:2010} and applied the heat source using the following boundary conditions 
\begin{equation}
\tau \frac{\partial T}{\partial t} = \frac{\pm2 \ell}{3} \frac{\partial T}{\partial x}, 
\end{equation}
where the positive and negative signs in the right hand side correspond to the $\xi=0$ and $\xi=1$ ends of the simulation cell, respectively. This condition is used to simulate a jump in the temperature in the boundaries when the steady state is reached as produced by ballistic phonons.

Figure \ref{Thermal-2} shows the time evolution of the temperature for the FE and the proposed model when Kn $=1$. The agreement between both methodologies is noteworthy and an indication that the proposed model is as good as the FE. We also compare the solution obtained in steady state for the EPRT \cite{Joshi:1993} and the BD \cite{Chen:2001} models. We see that the temperature profile is very close to the EPRT and the proposed model, but some differences appear {\color{blue} due to several reasons explained below. 

In both the BD and the EPRT models, the heat due to ballistic phonons is gradually introduced in the sample by using imposed fluxes to the heat carrier distributions \cite{Joshi:1993,Chen:2001}. On the other hand, in our model, we use initial temperature values and heat fluxes, which are difficult to link to the heat carrier distribution. Additionally, the BD and the EPRT suffer of an ill definition of temperature. For instance, in the BD there is no clear way to combine the ballistic and diffusive distributions to make a unique temperature. Thus, the BD results are rescaled to minimize the spurious defects of this ill definition.  Similarly, in the EPRT approach the phonon energy-flux distributions used to solve the BTE do not follow any equilibrium distribution form, making it difficult to obtain a temperature. To avoid this, Joshi \emph{et al.} \cite{Joshi:1993} computed the temperature from an equilibrium Bose-Einstein distribution that has the same average energy as the phonon energy-flux distributions of their solutions. Thus, a one-to-one comparison between models is difficult and leads to small discrepancies between models, as shown in Fig. \ref{Thermal-2}. Better agreement can be achieved if one tunes the how fast the heat flows into the system by adjusting the coefficients at the boundaries (see, for instance \cite{Alvarez:2010}). However, for the purposes of our comparison, we find that the boundary conditions used in our simulations give sufficient agreement with BD and EPRT.  }

We summarize this section by concluding that the proposed model is as good as the FE; our model can also account for size effects in devices by tuning the effective thermal conductivity of the device according to Eq. \ref{eq:Size-conductivity} provided the appropriate boundary conditions to emulate ballistic heat source generation; we can thus predict accurate steady states when the Knudsen number is close to one. 

\section{Thermo-mechanical coupling}\label{sec:ThermoMechanicalCoupling}

\begin{figure}
\centering
\subfloat[][]{\includegraphics[width=0.4\textwidth]{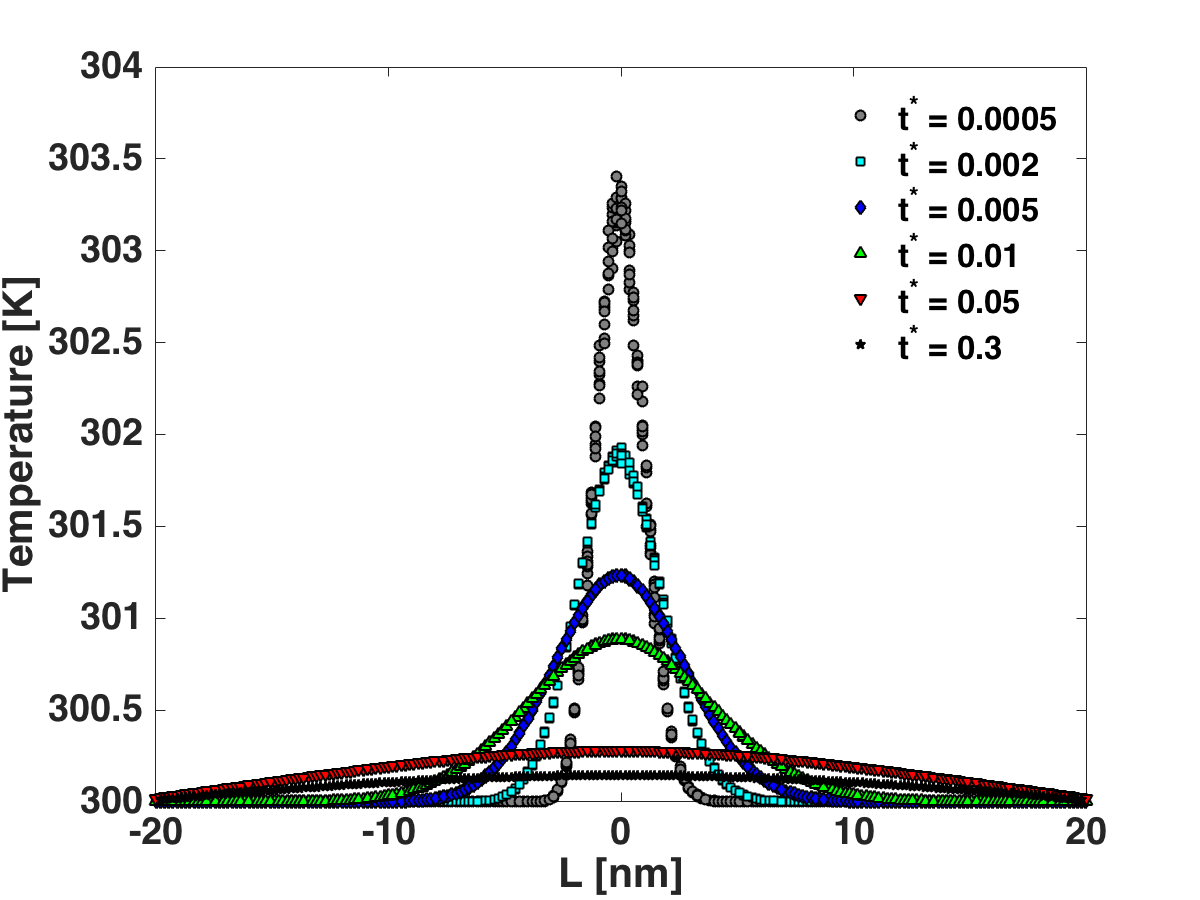}} 
\subfloat[][]{\includegraphics[width=0.15\textwidth]{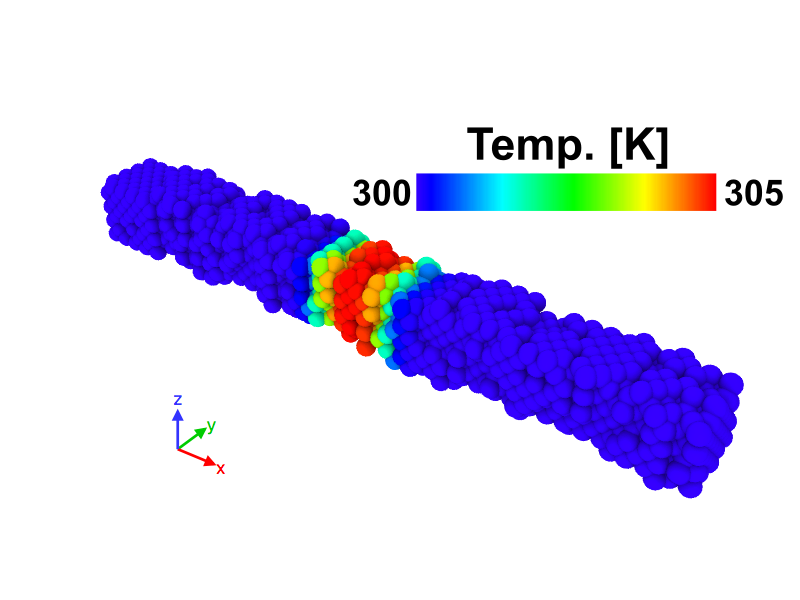}}
\subfloat[][]{\includegraphics[width=0.15\textwidth]{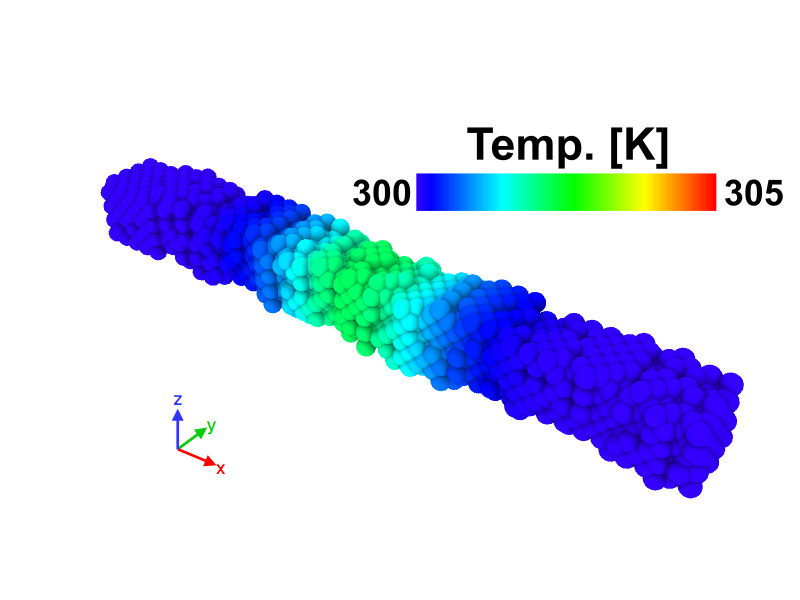}}
\subfloat[][]{\includegraphics[width=0.15\textwidth]{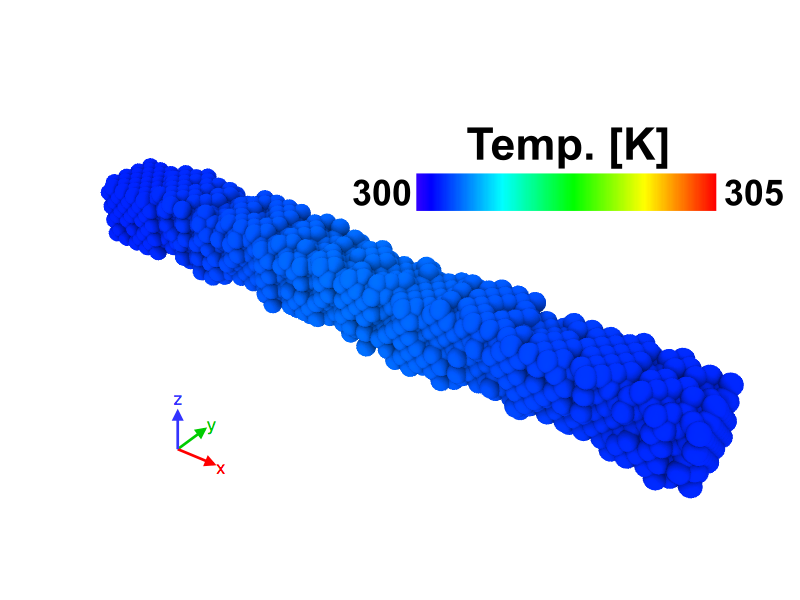}}
\caption{{\color[rgb]{0,0,1} a)} Evolution of temperature for each atom as a function of time using the proposed framework in a Cu bar with dimensions $l_x = 120 a_0 = 43.2$ nm, and $l_y = l_z = 4a_0 = 1.5$ nm and Kn = 0.58. Snapshots of the temperature field are taken at b) $t^* = 0$, c) $t^* = 5\times 10^{-4}$, and d) $t^* = 5\times 10^{-3}$, respectively. }
\label{Thermal-4}
\end{figure}

We now proceed to test the proposed \emph{unified} framework in a fully thermo-mechanical coupled formulation. In order to do so, we use the HotQC method \cite{Kulkarni:2008,Ariza:2012,Venturini:2014,Ponga:2015,Ponga:2016,Ponga:2017} to compute the free-energy of the system, Eq. \ref{Grand-Canonical-Free-Energy}, and update atomic temperatures and molar fractions using the transport laws given by Eqs. \ref{eq:HeatTransport} and \ref{eq:MassTransport}. We remark that in all examples, a fully atomistic resolution is retained and we do not introduce spatial coarse-graining. Moreover, the atomic interactions are taken into account by using suitable interatomic potentials. The equilibrium configuration are obtained in a variational way by minimizing the free-energy (Eq. \ref{Grand-Canonical-Free-Energy}) of the system with respect to atomic positions and vibrational atomic frequencies. We refer the reader to \cite{Kulkarni:2008,Ariza:2012,Venturini:2014,Ponga:2015} for a thorough treatment of the HotQC method and its implementation. 

\subsection{Heat conduction in a Cu bar} \label{Cu-bar}

We performed preliminary comparisons of our model's predictions to known material parameters, such as lattice parameters and stacking fault energies, for a variety of different materials at different temperatures. For the sake of brevity, we have omitted these results in this work and will show them in a forthcoming submission that will focus on the details of the implementation. Once we confirmed the validity of these results, we studied a thermo-mechanical coupled problem involving transfer of heat on a Cu bar of dimensions $l_x=120a_0=43.2$ nm, {\color{blue} $l_y = l_z = 4 a_0 \approx 1.5$ nm,} where $a_0 = 0.3615$ nm is the lattice parameter. {\color{blue}The mean free time was computed as $\tau = \frac{\ell}{v} = 5.25$ psec, with $\ell = 25$ nm and $v  =4.76$ nm/psec.} The bar was resolved at an atomistic level. For this application, we took Kn = 0.58 and the adjusted thermal conductivity using Eq. \ref{eq:Size-conductivity}, with $\lambda_0 = 405$ W/m $\cdot$ K. {\color{blue} This leads to an effective thermal conductivity of $\lambda = 170$ W/(m $\cdot$ K)}  We then applied an elevated heat pulse of $T_h=305$ K to a span of length $4a_0$ at the center of the simulation, holding the remainder of the bar at $T_c=300$ K, {\color{blue} i.e., $T(x = -21.5 \text{ nm},t) = T(x = 21.5 \text{ nm},t) = 300$ K; $T (|| x || < 0.7 \text{ nm}, t=0) = 305$ K.} 

The temperature profile is shown in Figure \ref{Thermal-4}. We see the outward diffusion of the heat pulse, ultimately resulting in thermal equilibrium of the bar. The behavior of this pulse coincides with predictions from FE heat flux using Eq. \ref{eq:Size-conductivity} for the thermal conductivity for a unidimensional system; this is an important result because of the fact that our formulation was able to bridge multiple length scales, providing details at atomistic length scales while matching continuum-scale predictions. More importantly, we see our model's ability to handle coupled thermo-mechanical problems with ease and accuracy.

\subsection{Thermal transport in a single-walled carbon nanotube}

\begin{figure}
\centering
\includegraphics[width=0.45\textwidth]{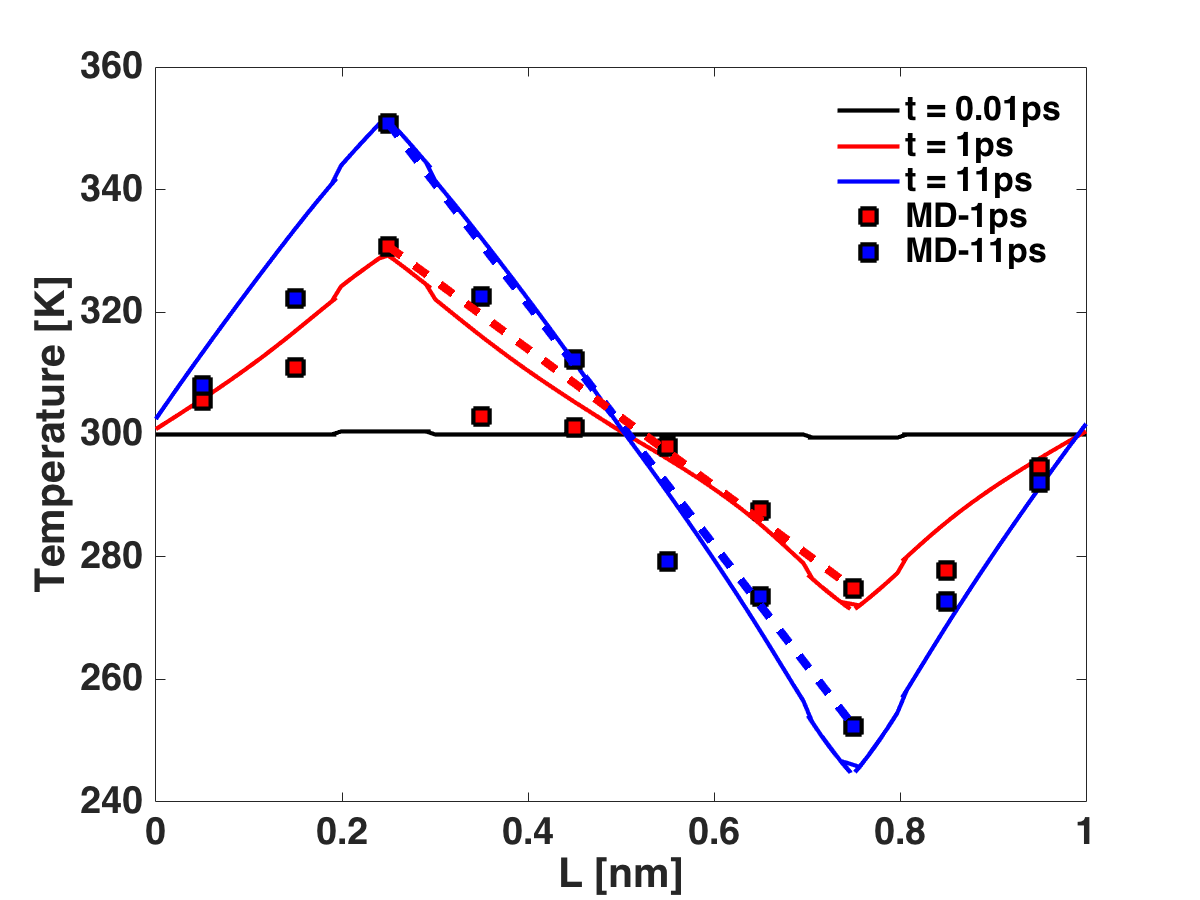}
\caption{Temperature evolution for a carbon nanotube with Neumann boundary conditions. A heat flux of $\pm 8$ eV/psec is applied to the atoms at two locations of the SWCNT. Square points show the NEMD result; solid lines show the result of the proposed model. Dashed lines is a linear interpolation for the NEMD points. }
\label{Thermal-6}
\end{figure}

To illustrate the ability of our framework to handle different materials and boundary conditions, we simulated {\color[rgb]{0,0,1} heat conduction} in a single-walled carbon nanotube (SWCNT) of length $L = 25$ nm and diameter $\phi = 1.4$ nm with a chiral index of $n=10$, $m=10$. The volume of the simulation cell was $V = 25 \times 2 \times 2 = 100$ nm$^3$, and this volume is used subsequently when virial stresses are reported. In order to simulate the atomic interactions, we used the REBO potential developed by  Brenner \emph{et al.} \cite{Brenner:2002}. We took $\lambda = 2980$ W/m$\cdot$K from previously-reported results \cite{Cagin:2000}, $v=20 $ nm/psec, $\ell \approx 500 $ nm, and $\tau = 25$ psec and computed $K_{ij}$ using these values. {\color{blue} In order to include ballistic effects, present in SWCNT of this size, we use a relaxation time $\tau_r = 1$ ps.} We used non-interacting boundary conditions in the $x-$ and $y-$ directions and periodic boundary conditions along the $z-$ direction.

In order understand the ability of the new model to predict transient fluxes, we endeavor to compare our results with non-equilibrium MD (NEMD) using the following conditions. We introduced a heat flux of $r_{in} = +0.02$ eV/(psec$\cdot$atom) to a set of atoms located between $0.2L$ and $0.3L$ and extracted the same amount, i.e., $r_{out} = -0.02$ eV/(psec$\cdot$atom) from a set of atoms between $0.7L$ and $0.8L$. The total number of atoms in each region was 400 atoms. Thus, the total heat flux injected/extracted was $q = 8$ eV/psec. The heat flux produced a temperature gradient in the sample, and this temperature gradient changed until it reached equilibrium after approximately 11 psec.  Figure \ref{Thermal-6}-a shows the temperature evolution in the SWCNT as a function of the time for both our new model and NEMD. Both the new model and NEMD reach approximately the same equilibrium temperature, and the profile of the atomic temperatures is approximately linear. For early times when the transient solution is present, we see that our model predicts a perfectly linear profile of temperatures, while NEMD (shown with square points in our plot) has some fluctuations. If one traces a line between the temperature location at $L=0.25$ and $L=0.75$, the temperature distribution between the two methodologies is very close. The red and blue dashed lines correspond to a linearly-interpolated profile obtained with NEMD at $t=1$ and $t=11$ psec, respectively. Due to the agreement with NEMD, our confidence in our model is increased.

{\color{blue}
\subsection{Anisotropic thermal conduction}  \label{AnisotropicModelValidation}
Let us now study the effect of anisotropic thermal conductivities in the sample. We took a simulation cell of Cu of dimensions $\l_x = 20 \text{ nm } \times \l_y = 30 \text{ nm }$, with an infinite $z-$ direction. Due to the different lengths of the sample, the thermal conductivity in the vertical and horizontal direction will be different, as indicated by Eq. \ref{eq:Size-conductivity}. In order to simulate this behavior, we took the parameters specified in Section \ref{Cu-bar}, leading to $r_{yx} = \frac{\alpha_y}{\alpha_x}= 1.65$, where $r_{yx}$ is the ratio between thermal diffusivities in the vertical and horizontal direction. In order to drive the system to a non-equilibrium thermodynamic state, we applied an initial heat pulse of $T_0(t=0) = 500$ K to the center of simulation cell. The initial heat pulse had a width of $0.5 \times 0.5$ nm$^2$. Then, the temperature of the atoms was allowed to change using the proposed heat conduction model. 

\begin{figure}
\centering
\subfloat[]{\includegraphics[width=0.225\textwidth]{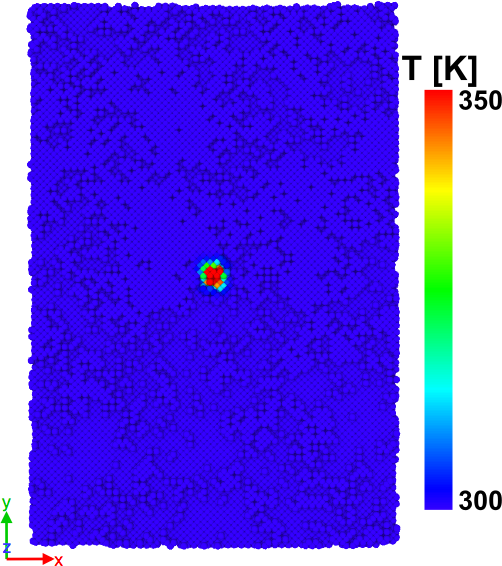}}
\subfloat[]{\includegraphics[width=0.225\textwidth]{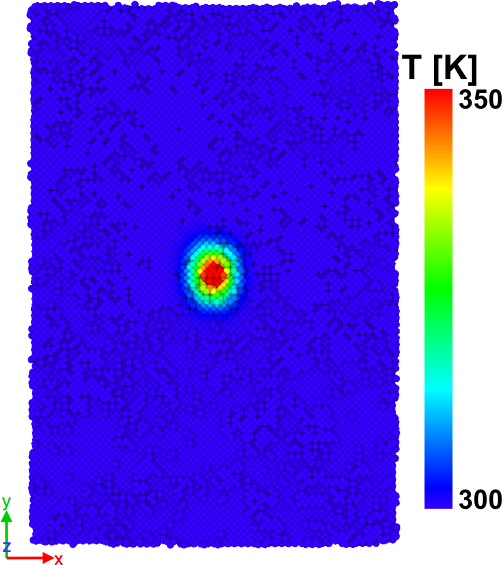}} 
\subfloat[]{\includegraphics[width=0.225\textwidth]{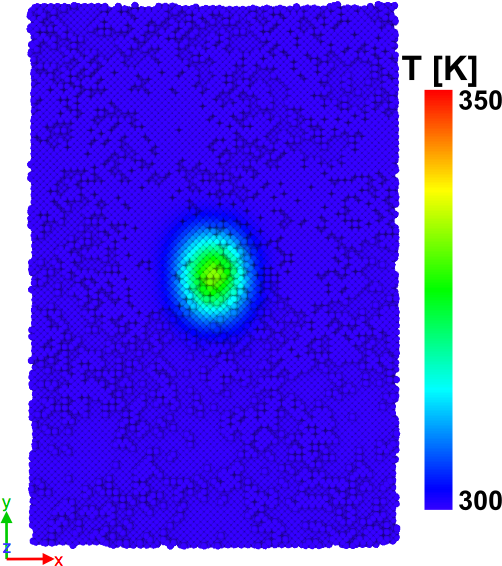}}
\subfloat[]{\includegraphics[width=0.225\textwidth]{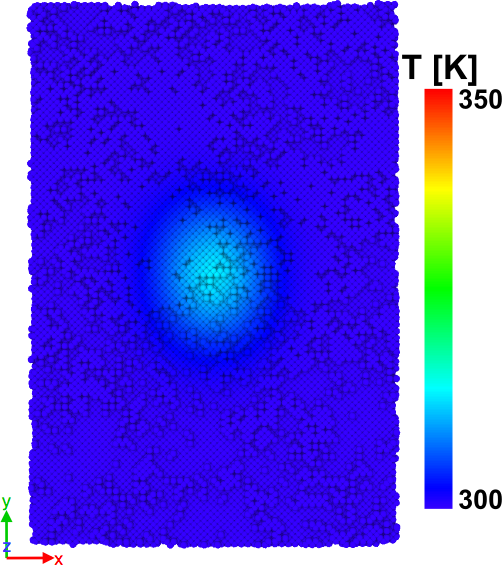}} 
\caption{Snapshots of the temperature distribution for a Cu sample with anisotropic thermal diffusivities with a ratio $r_{yx} = \frac{\alpha_y}{\alpha_x} = 1.65$. The temperature distribution correspond to a) $t^* = 0$, b) $t^* = 0.04$, c) $t^* = 0.2$, d) $t^* = 1.9$.}
\label{Anisotropic1}     
\end{figure}

Fig. \ref{Anisotropic1} shows four different snapshots of the temperature distribution in the sample from $t^* = 0$ to $t^* = 1.9$. The heat pulse diffuses as a function of time, increasing the temperature of nearby atoms. As expected,  the temperature distribution is anisotropic. In order to better understand the anisotropy, we analyze the temperature distribution for time $t^* = 0.04$ along the $x-$ and $y-$ directions and show the distribution in Fig. \ref{Anisotropic2}. We clearly see that the temperature profile along both directions can be fitted to a Gaussian function. The analytical solution for a heat pulse is a Gaussian profile, i.e., $T(t,x,y) \propto \exp{\left( -\frac{x^2}{4\pi \alpha_x t} \right)} \exp{\left( -\frac{y^2}{4\pi \alpha_y t} \right)}$. By adjusting the temperature distribution, we find that the ratio of the standard deviation for both directions is $r_{yx}^{\text{fitted}} = \alpha_y/\alpha_x = 1.52$, which is very close to the ratio of the thermal diffusivity.

\begin{figure}
\centering
\includegraphics[width=0.5\textwidth]{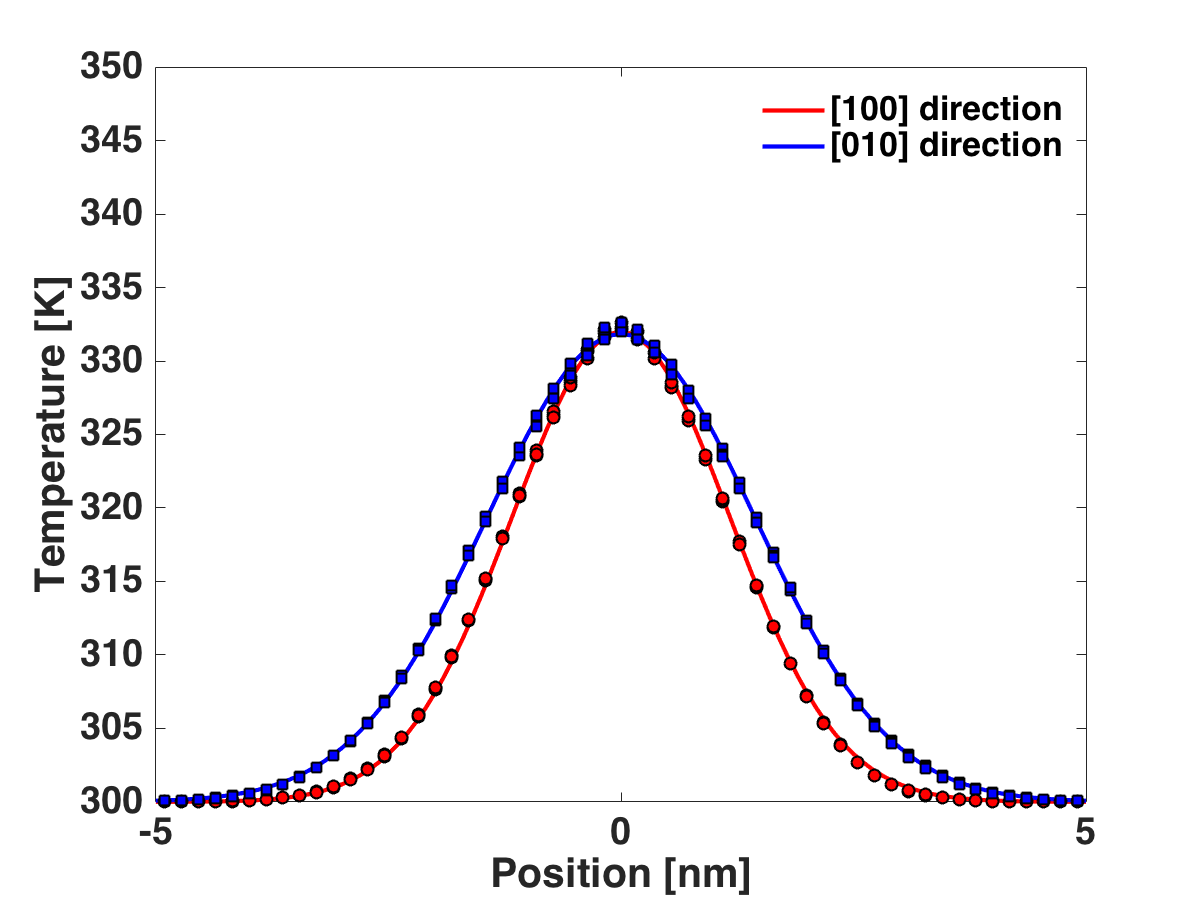}
\caption{Temperature distribution along the $[100]$ and $[010]$ directions for a plate of Cu at $t^* = 0.04$. The initial conditions correspond to a heat pulse with an initial temperature of $T_0(t=0) = 500$ K in the center of the sample. The spread of the temperature is larger in the vertical direction since the thermal conductivity is larger in that direction.  }
\label{Anisotropic2}
\end{figure}
}

\subsection{Thermo-buckling behavior of SWCNT under compression} 

\begin{figure}
\centering
\includegraphics[width=0.45\textwidth]{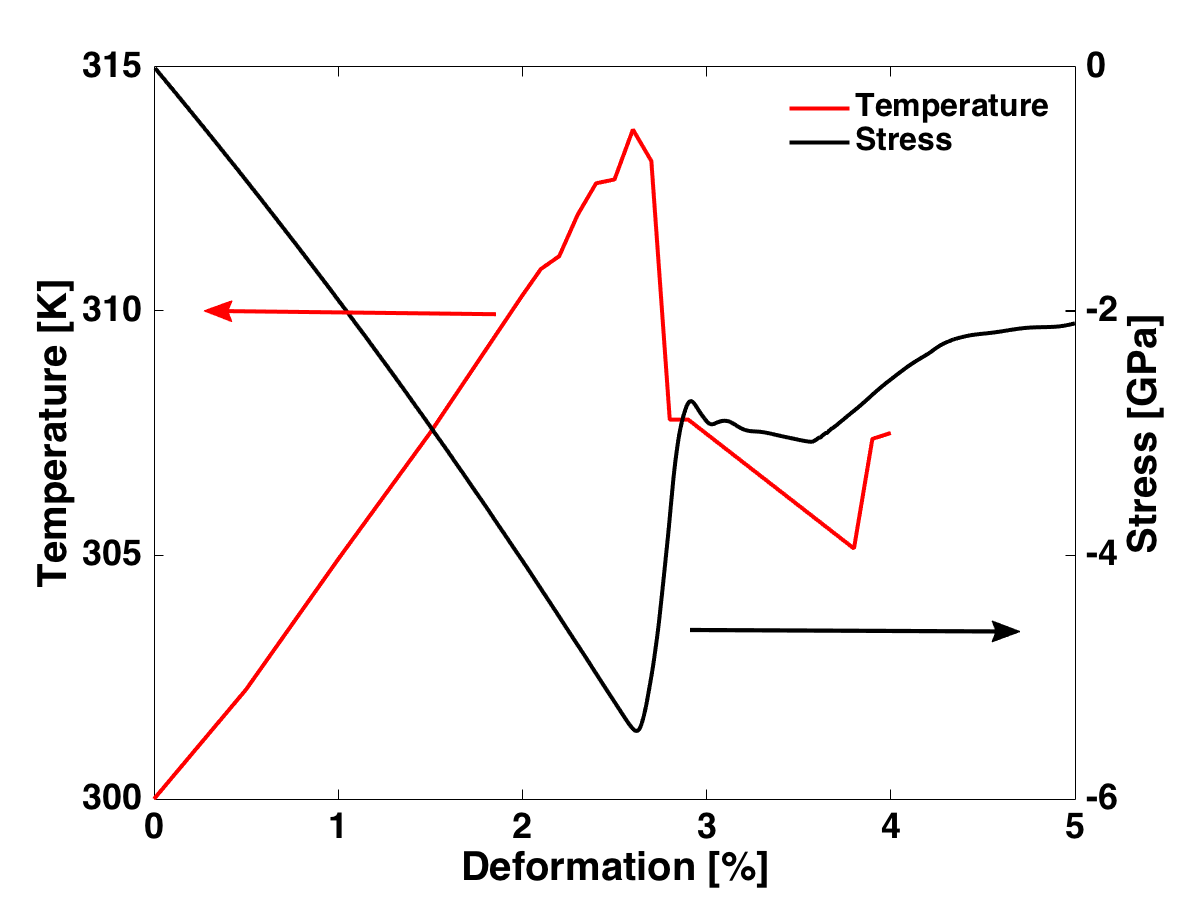}
\caption{Stress and temperature evolution as a function of the deformation when a SWCNT is subject to compression. A thermoelastic behavior is clearly evident in the simulation until the buckling of the SWCNT. Then, the temperature is highly non-linear and the values depend on the local environment. }
\label{CNT-thermomechanical}
\end{figure}

To illustrate the ability to predict the thermo-mechanical behavior of materials coupled with heat transport at the nanoscale, we applied a compressive load to the SWCNT described before and generated one vacancy in the middle of it. In order to simulate the coupled thermo-mechanical behavior of the SWCNT, we imposed an isentropic condition to the thermal vibration of atoms while loads were applied; this imposition was made such that when the local atomic frequencies change due to the collapse of the SWCNT, the temperature would then change in order to conserve the entropy of the system, i.e., $\frac{\omega_i^{n}}{T_i^{n}} = \frac{\omega_i^{n+1}}{T_i^{n+1}}$, where the superscripts $n$ and $n+1$ refer to a two consequent time steps. We initialized the temperature to $T=300$ K and coupled it with the new heat conduction model (Eq. \ref{eq:HeatTransport}) to regularize the temperature along the SWCNT. The time step used between different loading steps was taken to be $\Delta t = 2$ psec. This large time step was sufficient to homogenize the temperature field within 1K along the entire sample during the loading. Non-interaction boundary conditions in three directions were imposed in this simulation.

\begin{figure}
\centering
\subfloat[][]{\includegraphics[width=0.225\textwidth]{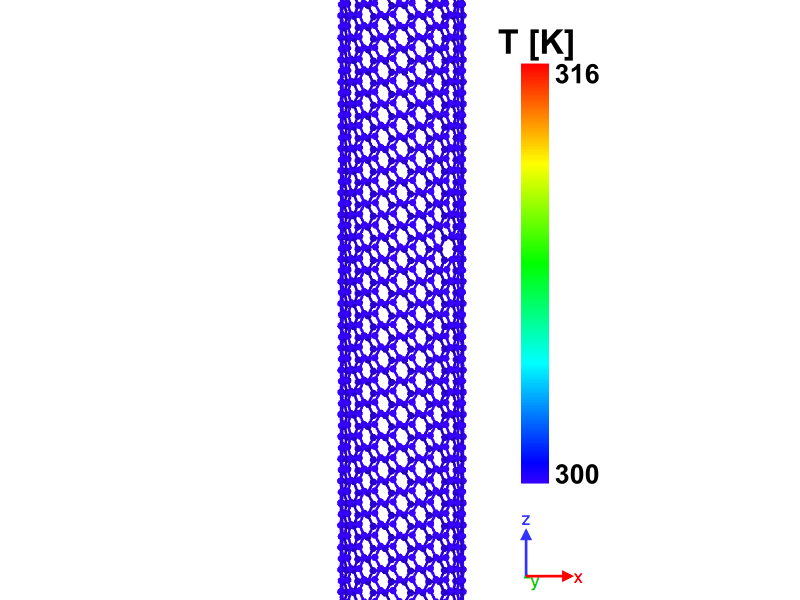}}
\subfloat[][]{\includegraphics[width=0.225\textwidth]{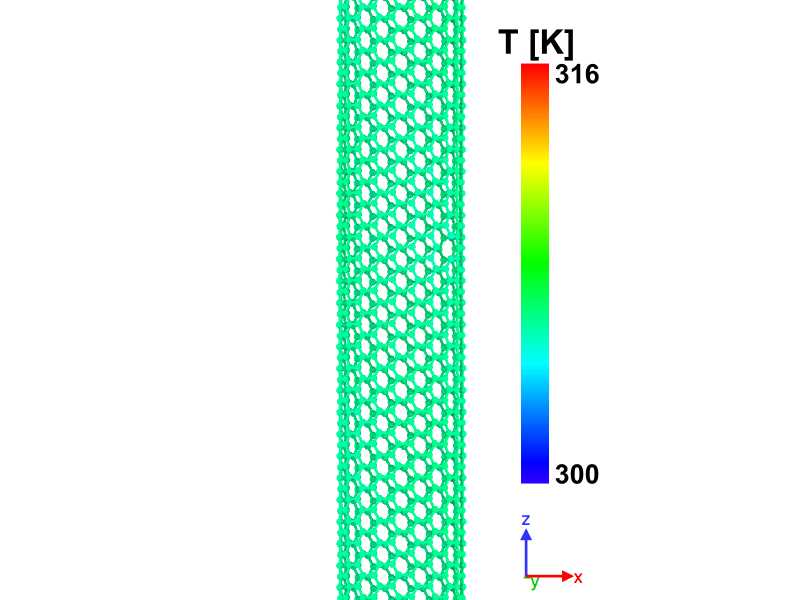}} 
\subfloat[][]{\includegraphics[width=0.225\textwidth]{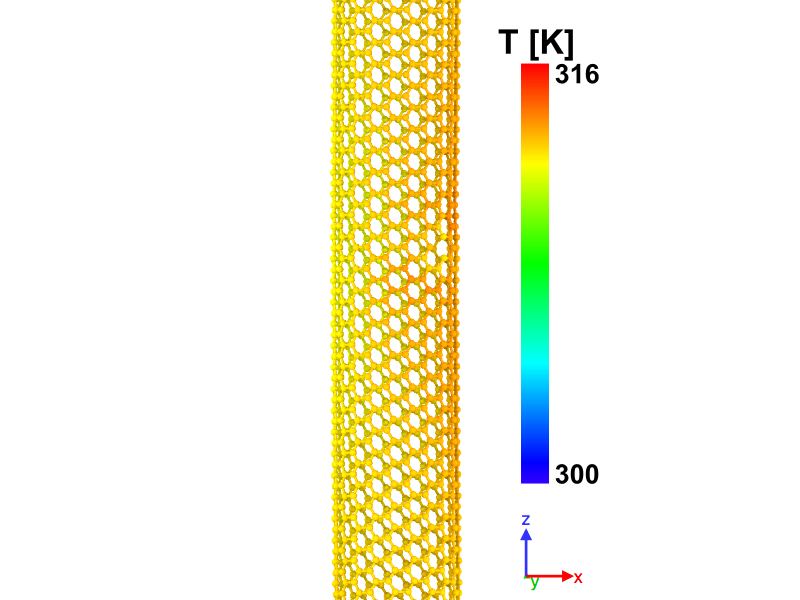}} \\
\subfloat[][]{\includegraphics[width=0.225\textwidth]{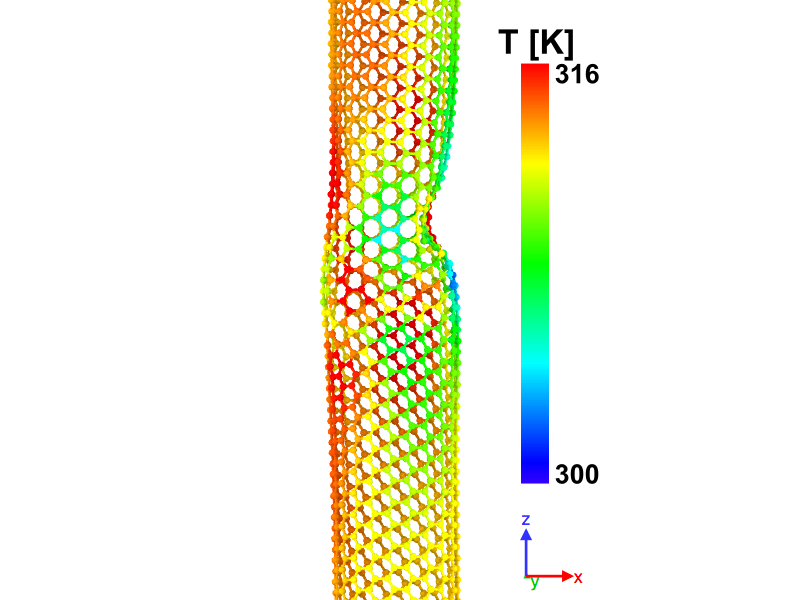}} 
\subfloat[][]{\includegraphics[width=0.225\textwidth]{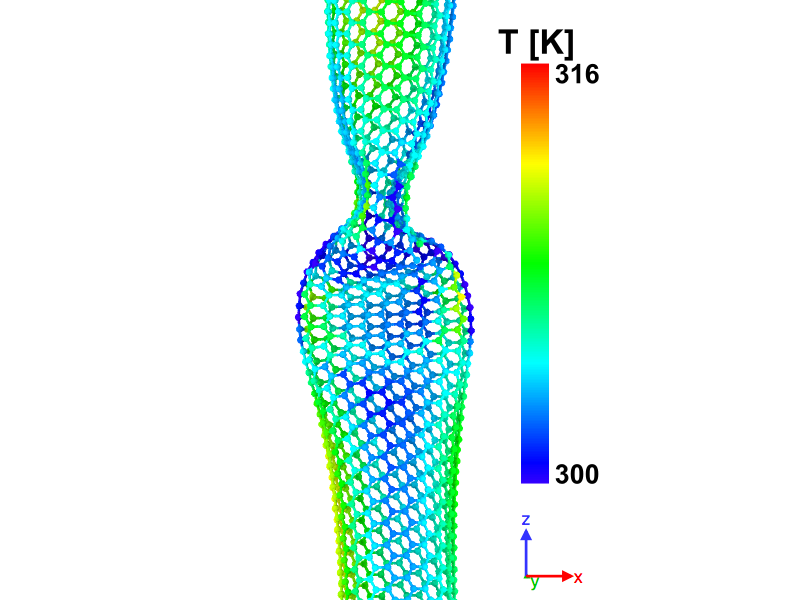}}
\subfloat[][]{\includegraphics[width=0.225\textwidth]{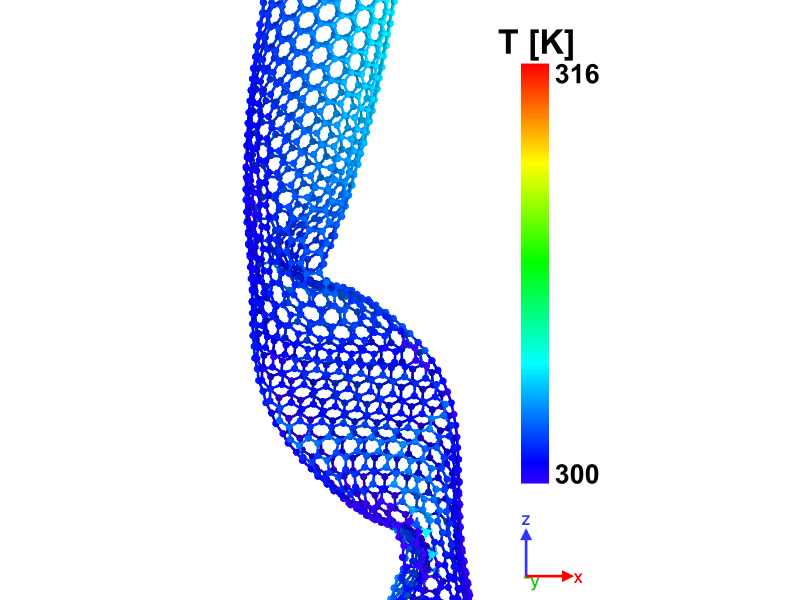}}
\caption{Deformation and temperature field experienced by the SWCNT under compression for different values of strain. a) Initial configuration, b) at $\epsilon = 1.25 \%$, c) at $\epsilon = 2.5 \%$, d) at $\epsilon = 2.6 \%$ e) at $\epsilon = 2.6 \%$ and f) at $\epsilon = 5.0 \%$. The carbon nanotube buckles after approximately at $\epsilon = 2.60 \%$ due to the applied compressive load. }
\label{CNT-thermomechanical-def}
\end{figure}

Figure \ref{CNT-thermomechanical} shows the thermo-mechanical behavior of SWCNT in compression. The stress vs. strain plot shows two regimes; an initial linear regime up to $\epsilon = 2.8 \%$ where the stress has reached a peak value of $\sigma_p = - 5.5$ GPa \footnote{The absolute values of the stress are irrelevant as it depends on the choice of the volume of the sample and this is arbitrary.}. Thereafter, we observed a deviation from the elastic solution, corresponding to a nonlinear buckling mode of the SWCNT causing a drop in the stress up to $\sigma = - 2$ GPa and a irreversible deformation of the SWCNT. The buckling instability is accompanied by a thermoplastic behavior in the SWCNT. During the linear regime, the temperature increases linearly up to the peak value and is homogeneous (Figures \ref{CNT-thermomechanical-def}-a to \ref{CNT-thermomechanical-def}-c). Once the SWCNT has buckled, the temperature drops and becomes highly heterogeneous (Figures \ref{CNT-thermomechanical-def}-d to \ref{CNT-thermomechanical-def}-f). In the later snapshots, individual values of temperature depend on the local environment of the atoms. This is due to the fact that, in order to generate the buckling instability, the SWCNT takes local kinetic energy from the temperature field and converts it to the macroscopic deformation and motion of the whole SWCNT that results in the buckling instability. Remarkably, the buckling motion has a much longer time scale than the phonon frequencies of the atoms; therefore, there is an energy exchange over different time scales that is linked with the heat conduction model. This illustrates our model's allowance for the direct coupling of the local atomic temperature with macroscopic kinetic motion of the atoms.

{\color{blue} Before closing this section, we provide some comments on the acceleration one can achieve by using the proposed framework for heat conduction at the nanoscale. We notice that the master equation is integrated with an implicit Euler algorithm, which is conditionally stable. In order to achieve convergence, the time step should  be less than $t_c = \frac{b^2}{\alpha}$. This leads to effective time steps that are of the order of a few femto-seconds. This is required since the characteristic time scale for heat transport is related to the phonon dynamics. However, the evaluation of the master equation is quite inexpensive in comparison with the evaluation of the potential, allowing one to perform multiple heat conduction steps between force evaluation. This allows for a time acceleration of the diffusive heat conduction phenomena, leading to time steps of approximately 2 psec or more, which is around 2000 times larger than the typical time step in MD. 
}

\section{Mass diffusion in nanoscale materials}

We now focus our attention on the ability of the proposed model to simulate mass transport in nanoscale devices and materials.  To illustrate this, we investigated the absorption and desorption of hydrogen (H) in palladium (Pd) nanospheres; this was chosen because of its strong relevance as an application for energy storage and because of the potential for direct comparisons to experiments performed in Langhammer \emph{et al.} \cite{Langhammer:2010}. In these experiments, a large surface effect appeared due to local atomic distortion of Pd atoms in these Pd-H alloys. The distortion then caused widespread rearrangement of Pd atoms, resulting in a diffusion of H that was considerably lower than bulk materials. We thus saw this as the perfect opportunity to apply our model, as the nature of the problem, which involves multiple, non-equilibrium nanoscale phenomena with large chemical gradients spanning multiple time scales, cannot be modeled with \emph{state-of-the-art} MD techniques.

\begin{figure}
\centering
\subfloat[]{\includegraphics[width=0.48\textwidth]{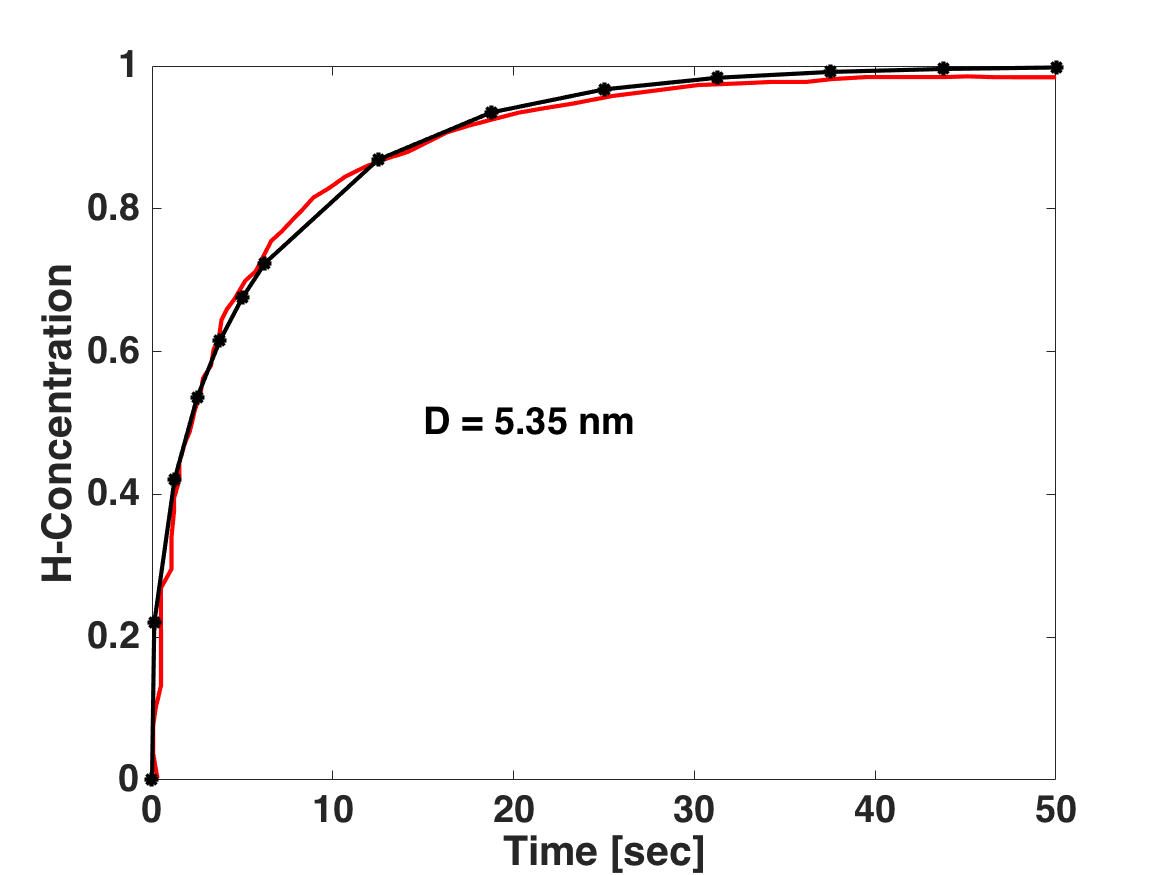}}
\subfloat[]{\includegraphics[width=0.48\textwidth]{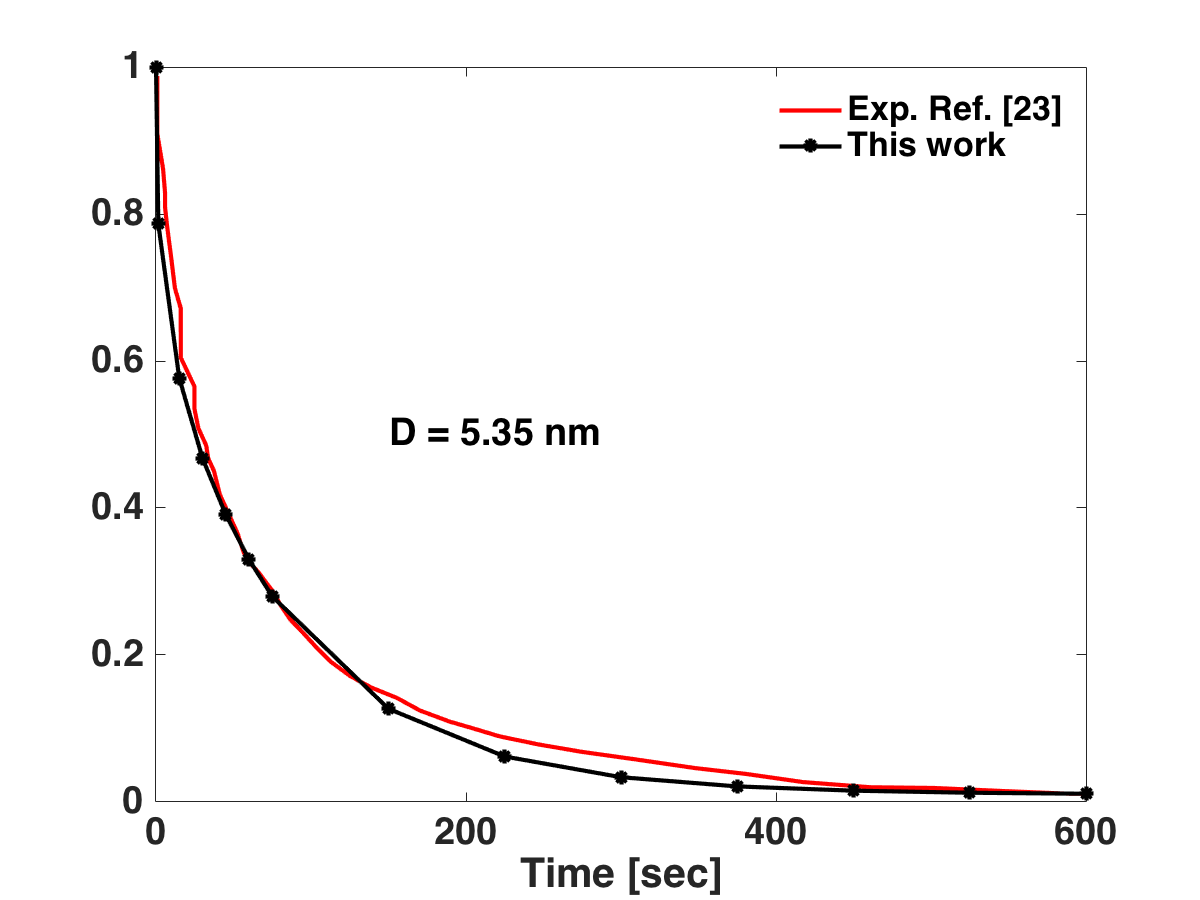}}
\caption{{\color[rgb]{0,0,1} Hydrogen absorption and desorption in Pd-nanospheres as obtained experimentally by Langhammer \emph{et al.} \cite{Langhammer:2010} (red line) and predicted with the proposed framework (black points). a) Absroption; b) Desorption.}}
\label{PdH-Diff-Nano}
\end{figure}

We chose a fixed temperature of $T = 330$ K and fixed the diffusivity of H in Pd to the experimental measures obtained by Langhammer \emph{et al.} \cite{Langhammer:2010}, using analytical solutions previously derived for spheres \cite{Crank:1979}. The system consisted of a Pd nanosphere of diameter $D = 5.35$ nm. The initial position of the Pd atoms was obtained by constructing a FCC lattice with $a_0 = 0.39$ nm. H sites were generated in the nanosphere by placing them in the octahedral interstitial {\color{blue} positions}. The Pd-H interactions were simulated using an interatomic potential developed by Zhou \emph{et al.} \cite{Zhou:2008}. After generating the initial positions, a relaxed configuration was obtained by minimizing the free-energy using a dynamic relaxation algorithm. The initial H concentration in all sites was set to be $x_H = 0.01$. 

In order to simulate H-absorption in the Pd nanosphere, a large concentration of H in the outermost layer of atoms was set up (in the desorption case, we switched the concentration of H and made it lower than what was on the nanosphere). The initially high/low concentration of H atoms in the outer layer of the nanosphere generated a sharp change in the chemical potential and therefore forced the diffusion of H towards the interior/exterior atoms. The chemical potential in the surface can be linked with the H pressure in the gas phase and can be directly linked to experimental pressures \cite{Chase:1998}. Our selections replicated the conditions when the H pressure was around $0.1$ atm for the charging process and $1\times 10^{-13}$ atm for the discharging process. 

After the first relaxed configuration was obtained, the atomic molar fraction of H-atoms was updated using Eq. \ref{eq:MassTransport}. This mass diffusion step generated a set of imbalanced forces; thus, a subsequent relaxation was performed. These steps were systematically repeated until the total H concentration converged to less than 0.01\% between iterations. We notice that the associated time scale is dominated by the diffusion of H into Pd. The critical time step is governed by a diffusive problem, i.e., Eq. \ref{eq:MassTransport} that is much larger than the characteristic phonon frequency of the atoms. This leads to time steps that are several order of magnitude larger than MD, {\color{blue} making it possible to reach simulation times that are comparable to experimental measures.}

Figure \ref{PdH-Diff-Nano}-a and b show the H concentration profile during charge/discharge, respectively. As we can see, our framework predicts a concentration profile that is in close agreement to the experimental values. While the H diffusivity was \emph{calibrated} to match the experimental data, the diffusion mechanisms and paths are dominated by changes in the \emph{local} chemical potential that is computed by using the free-energy of the system, Eq. \ref{Grand-Canonical-Free-Energy}. Thus, while the time scale is expected to match the experimental values, the actual concentration {\color{blue} values are} fully predicted by the framework.

\begin{figure}
\centering
\includegraphics[width=0.48\textwidth]{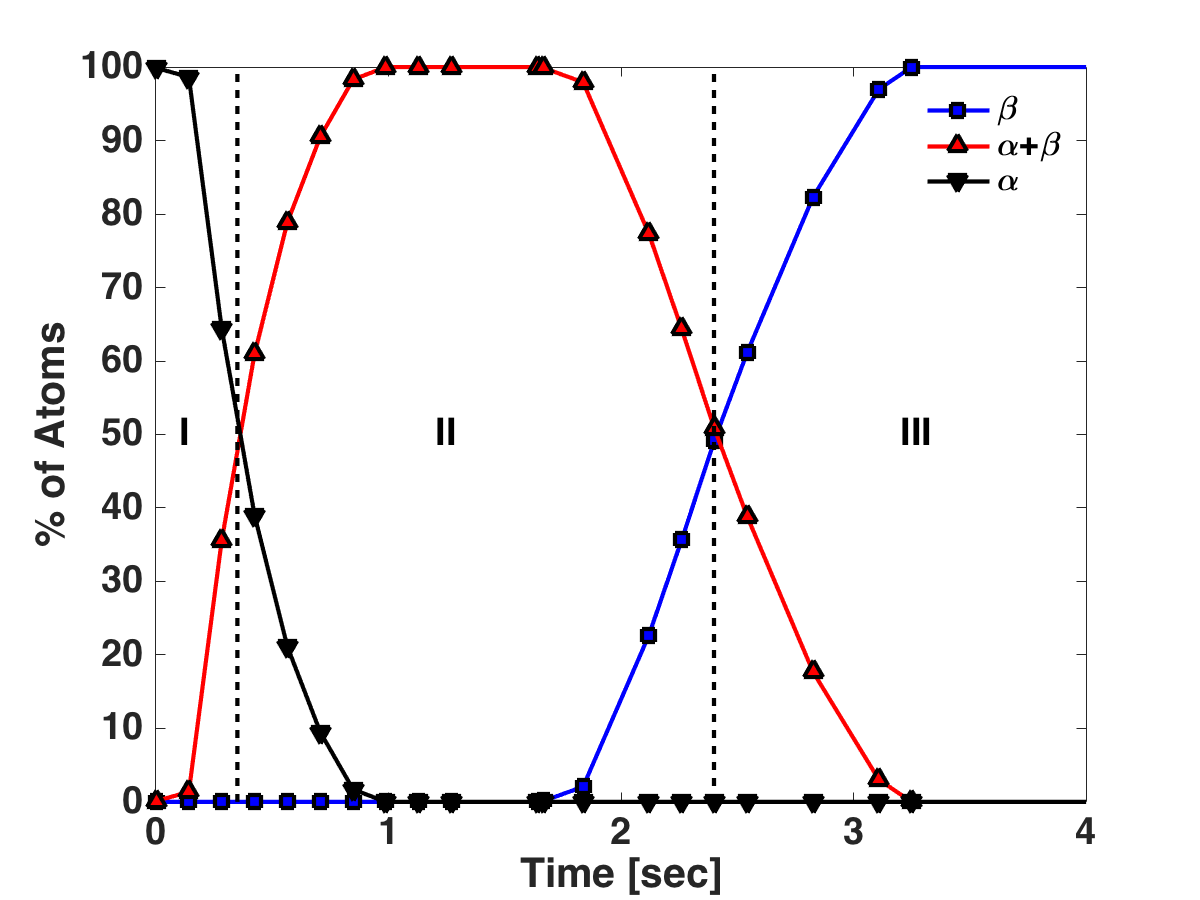}
\caption{Different phases during the charge of the nanosphere. Three different stages are observed which are characterized by atoms in the $\alpha$-phase, $\alpha+\beta$-phases, and $\beta$-phase. The three stages are delimited by the dashed vertical lines shown in the picture. }
\label{PdH-Phases}
\end{figure}

We then turn to understanding the mechanisms of diffusion in the Pd-H alloy. We select atoms with atomic molar fraction $x_H \le 0.15$ and label these as part of an $\alpha$ phase; we repeat this procedure, taking $x_H\ge 0.55$ for a $\beta$-phase, and labeling all the molar fractions in between as the $\alpha+\beta$-phase. This mixed region where $\alpha+\beta$-phase coexist is often called the miscibility gap. These threshold values were obtained from the work of Narehood \emph{et al.} \cite{Narehood:2009}, where X-ray diffraction patterns of small Pd particles were analyzed under different pressures and temperatures. Although there is a large dispersion in these values, we deem them sufficient for making a qualitative analysis of the phases. 

Fig. \ref{PdH-Phases} shows the distribution of atoms in the aforementioned phases. We immediately notice that the phases are characterized by smooth boundaries, which is an indication of the overlap of the phases during the charge. Examining the start and end times of when the percentage of in-phase atoms exceeds 50\% in each phase, we see that stage I --- corresponding to low H concentrations ($\alpha$-phase) --- exists only for a brief period of time of approximately 0.35 sec. The mixed $\alpha+\beta$-phase, represented by stage II, extends from 0.35 to 2.4 sec. Thereafter, we observe a third stage that holds for the remainder of the simulation. This allows us to see the transient nature of the noncrystalline states prior to the end result of the atoms sitting in octahedral sites. 

\begin{figure}
\centering
\includegraphics[width=0.48\textwidth]{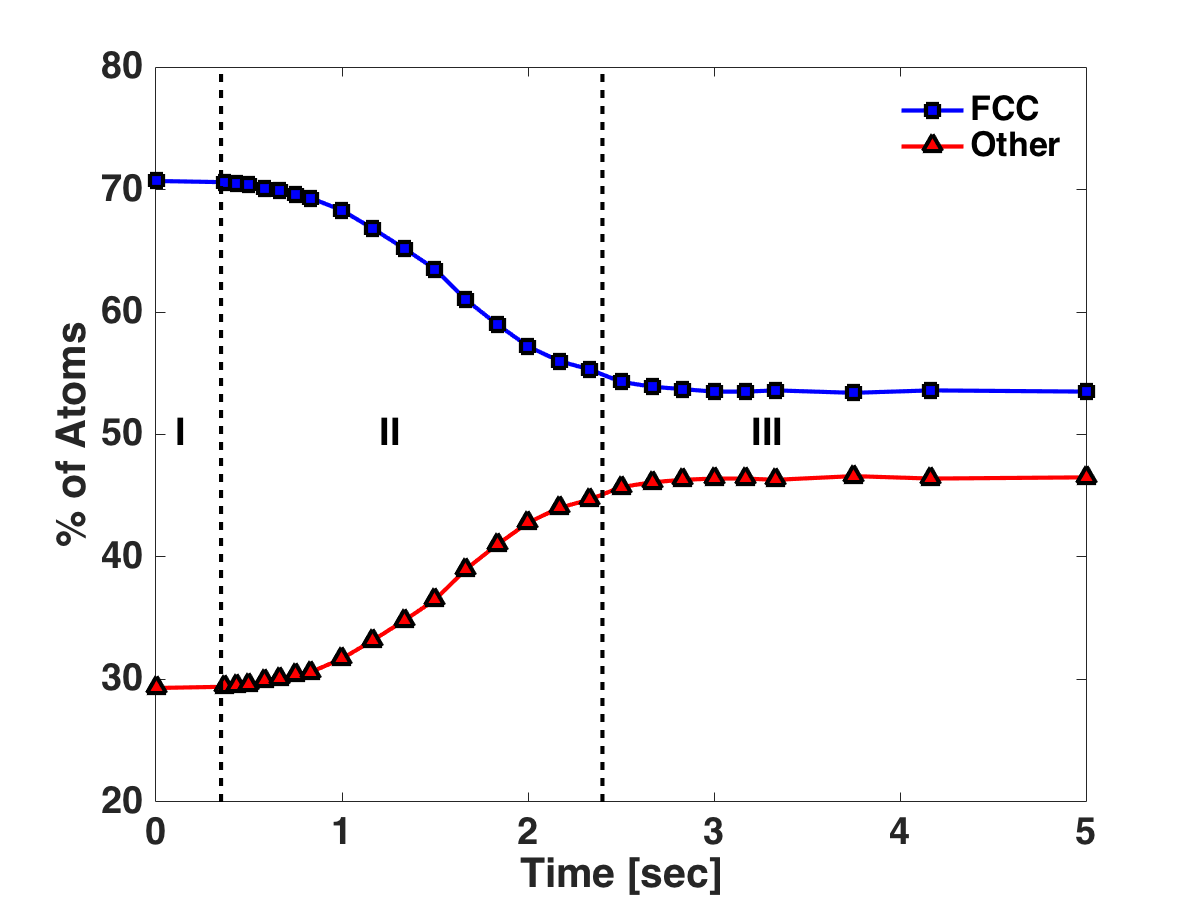}
\caption{Different phases during the charge of the nanosphere and local atomic structure of Pd atoms during the charge. Vertical dashed lines indicate the three different phases observed during the simulation.}
\label{PdH-Structure}
\end{figure}
Finally, we examine the surface effects driven by H absorption during charging. In order to understand the surface effects from a qualitative point of view, we investigated the local atomic structure of Pd atoms during the charge. Using Common Neighbor Analysis (CNA) \cite{Stukowski:2012}, we distributed the atoms into those with an FCC structure and those in an amorphous state. The results are plotted in Fig. \ref{PdH-Structure}. In the initial state, approximately 70\% of the atoms are in an FCC structure, while the remaining 30\% are in an amorphous state because they are on the surface of the nanosphere. However, by the end of the simulation, the distribution of atoms in the FCC structure has reduced to 55\%. This suggests that there is a large atomic relaxation in the surface due to the absorption of H atoms.

Moreover, our simulations predict a layer of amorphous Pd atoms of approximately 0.5 nm thickness in the surface of the nanosphere. This amorphous layer distorts the atoms and obstructs the H diffusion towards the interior of the nanosphere resulting in a lower diffusivity than macroscopic specimens \cite{Langhammer:2010,Narayan:2017}. The local structure of the nanosphere in the interior of the shell, however, remains FCC. Of note, when the H is fully removed during the desorption process, 65\% of the Pd atoms remained in an FCC structure and this observation seems to be in agreement with experimental observations \cite{Narehood:2009,Narayan:2017}. 

This investigation of H diffusion in Pd nanosphere is a good demonstration of a non-equilibrium phenomenon that involves long time scales with large chemical potential gradients. Our methodology allows for the simultaneous thermo-chemo-mechanical coupling over multiple time scales that would be impossible to simulate with the state-of-the-art MD techniques. 

{\color{blue} We now comment on the computational cost of the implementation and the speed-up that is achieved with respect to MD. Due to the ability of the framework to compute free-energies and chemical potentials that are functions of the atomic molar fractions, is it possible to use a kinematic diffusion law given by the \emph{master equation} to simulate diffusive phenomena. This combination eschews the need to simulate individual atomic hops from one site to another, which otherwise would take an exceedingly large amount of time in traditional MD. Considering the fact that, in our framework the full absorption is achieved after only 30,000 of force evaluations, the speed-up obtained is estimated to be at least of the order of $10^7$ with respect to MD. This acceleration is enormous and illustrates the need to develop frameworks such as the one introduced in this paper in order to achieve realistic time scales in atomic scale simulations. Finally, we comment on the computational cost of the simulations. The absorption/desorption simulations were carried out in a linux machine with four Intel$^\circledR$ Core$^\circledR$ i5-2439 @2.40 GHz processors. The total time for completion was around five hours.  Additional details of the formulation and further extensions and comparisons examples illustrating the potential of our implementation will be shown in our forthcoming work \cite{PongaSun2017b_HeatTransport}.}

\section{Conclusions}

In summary, we have developed a new model for heat conduction at the nanoscale. In contrast with the classical Fourier model, where heat is driven by temperature gradients, our proposed model uses atomic level information to predict effective energy exchange rates between neighboring sites. The proposed model for heat diffusion shares the same structure as the governing equation for mass transport, providing a unified framework for simulating heat and mass transport at the nanoscale. The model is consistent with the first and second laws of thermodynamics and does not require the use of a basis set to evaluate operators. This conveniently allows for seamless implementation in atomic models without the need to evaluate operators using basis sets. Remarkably, the model correctly reproduces classical results when Kn $\ll 1$.  Moreover, when fitted to experimentally-measured thermal conductivity of small devices, the model is capable of predicting steady state behavior in scenarios where Kn is close to $ 1$.

We have validated our framework with several previously-developed models for heat conduction including NEMD and BTE approaches, spanning a wide range of Kn. In all validation examples, our model produced very accurate transient profiles, which were directly comparable with the FE. We have also coupled the model with a non-equilibrium thermodynamic framework to simulate thermo-chemo-mechanical problems at the nanoscale. Through multiple examples, we have shown that our framework is an extremely powerful simulation tool, capable of resolving many different complicated transport problems at an atomistic scale while also providing valuable information which can be validated directly with experimental measures.

{\color[rgb]{0,0,1} It should be mentioned that our model is not without its limitations. In particular, our model has not been tailored for small systems where quantum effects, such as quantum confinement, are present. Here, quantum effects would be much prevalent in determining the behavior of the system. Capturing such behavior is extremely difficult and would require further development of our model and might be interesting in the spirit of continued bridging of length scales.}

{\color[rgb]{0,0,1} Nevertheless, for most devices at the atomic scale with reasonable system sizes, we find our model to provide novel insights that would not have been possible with previously state-of-the-art simulation techniques.}  We also point out that our proposed model for heat conduction can be applied to a variety of particle-based methodologies that lack of basis set {\color{blue} and can be also used to simulate electronic temperature with traditional MD methods within the context of two-temperature models}. These are avenues that the authors are actively pursuing.  

\section{Acknowledgments}
We gratefully acknowledge the support from the Natural Sciences and Engineering Research Council of Canada (NSERC) through the Discovery Grant under Award Application Number RGPIN-2016-06114 and the support of Compute Canada through the Westgrid consortium. This research used resources of the Argonne Leadership Computing Facility, which is a DOE Office of Science User Facility supported under Contract DE-AC02-06CH11357.

\section*{References}
\bibliography{myreferences}

\end{document}